\documentclass[a4paper,11pt]{article}
\pdfoutput=1 

\usepackage{jheppub_modified} 

\usepackage[T1]{fontenc} 

\usepackage{verbatim}
\usepackage{color}
\usepackage{graphicx}
\usepackage{subcaption}

\usepackage{sectsty}
\allsectionsfont{\boldmath}

\newcommand{\eq}[1]{Eq.~(\ref{#1})}
\newcommand{\bib}[1]{Ref.~\cite{#1}}
\newcommand{\refs}[1]{Refs.~\cite{#1}}
\newcommand{\bibs}[1]{\cite{#1}}
\newcommand{\fig}[1]{Fig.~\ref{#1}}
\newcommand{\tab}[1]{Table~\ref{#1}}

\newcommand{\sect}[1]{Section~\ref{#1}}

\newcommand{\appen}[1]{Appendix~\ref{#1}}

\def\Ref#1{Ref.~\cite{#1}}

\newcommand{\bea}{\begin{eqnarray}}
\newcommand{\eea}{\end{eqnarray}}

\def\MG5{{\tt MadGraph5\_aMC@NLO}}

\newcommand{\gev}{{\unskip\,\text{GeV}}}
\newcommand{\tev}{{\unskip\,\text{TeV}}}

\title{Next-to-leading order QCD corrections for single top-quark 
production in association with two jets}

\author[a]{Stefan M{\"o}lbitz,}
\author[b]{Le Duc Ninh}
\author[a]{Peter Uwer}

\affiliation[a]{Humboldt-Universit\"at zu Berlin, Institut f\"ur Physik,\\
Newtonstra{\ss}e 15, D-12489 Berlin, Germany}
\affiliation[b]{Institute For Interdisciplinary Research in Science and Education,\\  
ICISE, 590000 Quy Nhon, Vietnam}

\emailAdd{moelbitz@physik.hu-berlin.de}
\emailAdd{ldninh@ifirse.icise.vn}
\emailAdd{Peter.Uwer@physik.hu-berlin.de}

\preprint{HU-EP-19/16, IFIRSE-TH-2019-3}      
\abstract{In this article we calculate the next-to-leading order (NLO)
  QCD corrections for single on-shell top-quark production in
  association with two jets at proton-proton colliders. The $tW$
  channel is assumed to be measured independently. The QCD corrections
  to the inclusive cross section are about 28 (22)\% for top
  (anti-top) quark production at the 13 TeV LHC. Theoretical errors
  are dominated by scale uncertainties, which are found to be around
  5\% at NLO. Results for various kinematical distributions are also
  provided using a well-motivated dynamical scale. The QCD corrections
  are found to have a non-trivial dependence on the phase-space.  }

\begin{document}
\maketitle
\flushbottom

\section{Introduction}
\label{sect:intro}
Twenty-five years after its discovery many questions related to the top
quark are still open, despite the tremendous progress made in recent
years concerning the measurement of its mass and its interactions. 
Why is the top-quark the only quark with a natural
Yukawa coupling to the Higgs boson of order one ? Why is it almost 35
times heavier than the next heavy quark, the $b$-quark ? Are the
top-quark's weak interactions as in the Standard Model or does the
top-quark play a special r\^ole in the electroweak symmetry breaking
as predicted in many extensions of the Standard Model ? 

The hadronic production of single top quarks allows to shed light on
these questions. In particular, singly produced top-quarks provide an
ideal laboratory to study the top-quark weak interactions. This is a
major difference to top-quark pair production---with a roughly three
times larger cross section, the dominant process for top-quark
production in hadronic collisions---where the weak couplings are only
accessible through the top-quark decay. Since most of the experimental
effort in recent years was devoted to pair production, the top quark's
weak interactions are currently experimentally much less constrained
through direct measurements than the top-quark strong interactions.  A
detailed study of single top-quark production offers the opportunity
to fill this gap and to search for new physics.  Furthermore, single
top-quark production provides complementary information compared to
top-quark pair production and allows studies not possible in top-quark
pair production. While top quarks produced in pairs are to good
approximation unpolarized (a tiny polarization is generated by QCD
absorptive parts and weak corrections \cite{Bernreuther:1995cx,
  Dharmaratna:1989jr,Bernreuther:2010ny}), singly produced top quarks
are highly polarized. Single top-quark production presents thus a
unique source of polarized top quarks which can be used for detailed
tests of the $V-A$ structure of the coupling to the $W$ boson and to
constrain potential new physics. In addition, single top-quark
production offers a direct handle to measure the
Cabibbo-Kobayashi-Maskawa (CKM) matrix element $V_{tb}$---providing
complementary information to indirect determinations based on
unitarity.

Single top-quark production was first observed in proton--anti-proton
collisions at the Tevatron \cite{Aaltonen:2009jj,Abazov:2009ii}.
According to the virtuality of the $W$ boson occurring in the Born
approximation, three different channels are distinguished: $s$-channel
production with $p_w^2>0$ ($p_w$ denotes the four momentum of the
$W$ boson), $t$-channel production with $p_w^2<0$, and the $tW$
channel where the $W$ boson occurs in the final state.  The
$t$-channel production is the dominant production process at the
Tevatron and the LHC. At the Tevatron $s$-channel production is the
second important channel.  The $tW$ channel is suppressed at the
Tevatron because of the limited collider energy of only 1.96 TeV. At
the LHC the situation is reversed. The $tW$ channel represents the
second important channel while the $s$-channel production is
suppressed. Because of the challenging experimental environment and
large backgrounds so far only evidence for $s$-channel production has been
reported by the ATLAS experiment \cite{Aad:2015upn}.

For all three production channels the next-to-leading order (NLO) QCD
corrections have been calculated
\cite{Bordes:1994ki,Stelzer:1997ns,Stelzer:1998ni,Harris:2002md,
  Sullivan:2004ie,Sullivan:2005ar,Giele:1995kr,Zhu:2002uj,Smith:1996ij}.
While initially only the inclusive cross sections have been analyzed,
later works include also results for differential cross sections. In
addition, the effects of the top-quark decay and the parton shower
were analyzed
\cite{Campbell:2004ch,Cao:2004ky,Cao:2005pq,Frixione:2005vw,
  Campbell:2005bb,Frixione:2008yi,Re:2010bp,Cao:2004ap}.
Based on soft gluon resummation, approximate next-to-next-to-leading
order (NNLO) results have been published in
\refs{Mrenna:1997wp,Zhu:2002uj,Kidonakis:2006bu,Kidonakis:2007ej,%
  Kidonakis:2010ux,%
  Kidonakis:2010dk,Kidonakis:2011wy}. An important step towards full
NNLO results has been made for $t$-channel production in
\refs{Brucherseifer:2014ama,Berger:2016oht} where NNLO results within
the leading-color approximation are presented.  Restricting the
analysis to the leading-color contribution, the calculation of the NNLO
corrections is significantly simplified, since the double box
contributions, notoriously difficult to calculate, drop out. As a step
beyond this approximation the reduction of the double-box topologies
to master integrals has been performed in
\cite{Assadsolimani:2014oga}. However, the occurring master integrals
are still unknown, although progress towards their evaluation has been
made in \cite{Meyer:2016slj}, where some of the integrals are studied
as sample applications.  Recently, the studies within the leading
color approximation have been extended to include also the top-quark
decay allowing to study single top-quark production fully
differentially at the level of the decay products
\cite{Berger:2017zof}. 

Already at next-to-leading order, real
corrections with an additional jet in the final state start to
contribute. In fact, a detailed study shows that a significant
fraction of single top-quark events is produced with additional jet
activity. Demanding a minimal $p_\perp$ of 25 GeV, about 30 \% of
singly produced top-quark events are produced in association with two
jets. To make optimal use of the data collected at the LHC, precise
predictions for single top-quark production in association with two
jets are mandatory. For reliable theory predictions at least NLO QCD
corrections are required. Furthermore, the NLO QCD corrections to
single top-quark production in association with two jets contribute to
single top-quark production at NNLO QCD and are thus required to
extend the existing leading-color results. In this article, we present
the NLO corrections to single top-quark production in association with
two additional light jets. In principle, the NLO corrections can be
produced with publicly available tools like for example
\MG5~\cite{MG5}
or {\tt GoSam}
\cite{Cullen:2014yla}. However, similar to single top-quark production
the $t$-channel production needs to be separated from the $tW$ production,
which requires to remove some of the Feynman diagrams contributing to the
full amplitude. Furthermore, the NLO corrections to single top-quark
production in association with two additional jets contribute to the
NNLO corrections to the inclusive single top-quark production. With this
application in mind, where a highly optimized execution might be
crucial, we have decided to do the calculation by a direct evaluation of
the Feynman diagrams and use {\tt GoSam} only to partially cross check
the results.

Let us mention that during the work on this project, similar results
have been published in \Ref{Carrazza:2018mix}. In this article, 
the authors work in the leading color
approximation which is for the concrete process equivalent to the
so-called structure function approximation. The key ingredient is
that the QCD corrections are studied independently for the two
incoming quark lines. At the same time this approximation gives also a
clear separation from other single top-quark processes since
interference terms are color suppressed. In addition to the fixed
order calculation within the leading-color approximation, the results
are further improved using the MINLO approach
\cite{Hamilton:2012np}. We consider the results presented in
  \Ref{Carrazza:2018mix} as complementary to the ones presented here.

The outline of the article is as follows. In \sect{sect:cal} we
summarize the calculation and present some technical details. In
\sect{sect:results} we describe the numerical input 
and present results for inclusive cross sections and various kinematical 
distributions. In addition we discuss the main uncertainties. We
present a detailed discussion of the scale uncertainties and show the
improvements using a dynamical scale. In \sect{sect:conclusion} 
conclusions are given. 
In \appen{appen:additional_plots} we show additional results for the production
of anti top quarks.

\section{Calculation}
\label{sect:cal}
We consider on-shell production of a single top quark in association
with two jets in proton-proton collisions. We work in the five flavor
scheme. The bottom quark is thus treated as massless  and
considered as part of the proton. We neglect the generation mixing terms
in the CKM matrix since these contributions are further suppressed by
the parton distribution functions. Employing the unitarity of the CKM
matrix this approximation is equivalent to replace the CKM matrix by
the identity matrix. Using the identity matrix for the CKM matrix
leads to a CP invariant theory. It is thus sufficient to study
top-quark production since the results for anti--top-quark production
can be obtained from CP invariance.

Care must be taken to separate top-quark production in association
with two jets from from the $tW$ channel with subsequent hadronic
decay of the $W$-boson. The latter process leads to the same final state
but is measured separately by the experiments. This issue is well
known from single top-quark production. In leading-order the
interference term between the two contributions vanishes and the
individual contributions are gauge invariant. We assume that these
contributions are small in next-to-leading order---in particular, when
experimental cuts to separate the $tW$ channel are applied.  For a
similar discussion we refer to \Ref{Carrazza:2018mix} where this
approximation is used to justify the structure function
approximation. Whenever we refer to the $tjj$ channel in the
following, the contribution from on-shell $tW$ production and
subsequent decay is removed.  More details will be given below.

\subsection{Leading order}
\label{sect:LO}
At LO, all subprocesses can be classified into two groups: with or
without $\bar{U}_iD_i$ in the final state, where $i=1,2$ for the first
two generations and capital letters are used to denote the generic up-
or down-type light quarks.  All subprocesses without $\bar{U}_iD_i$ in
the final state belong to the $tjj$ channel.

\begin{figure}
\begin{center}
\begin{subfigure}{0.30\textwidth}
  \includegraphics[width=0.8\textwidth]{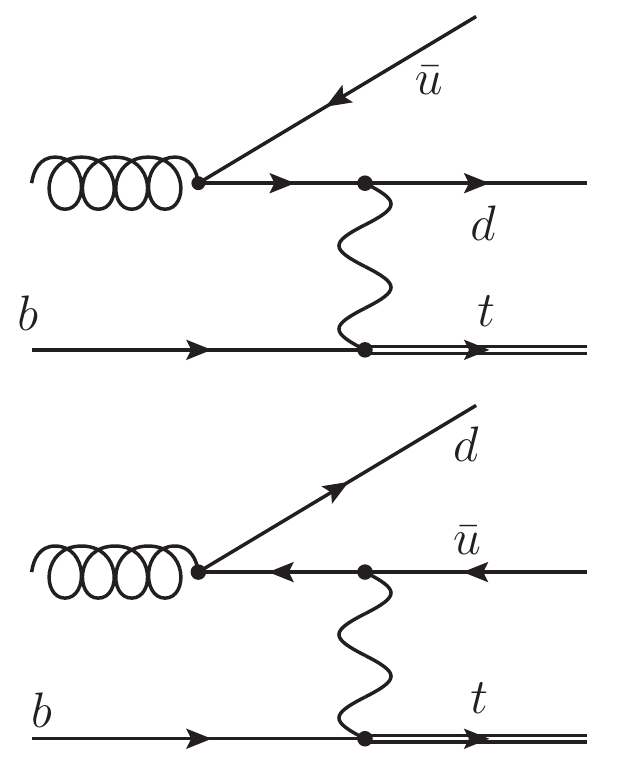}
  \caption{$tjj$}
  \label{fig:gq1}
\end{subfigure}
\begin{subfigure}{0.30\textwidth}
  \includegraphics[width=1.0\textwidth]{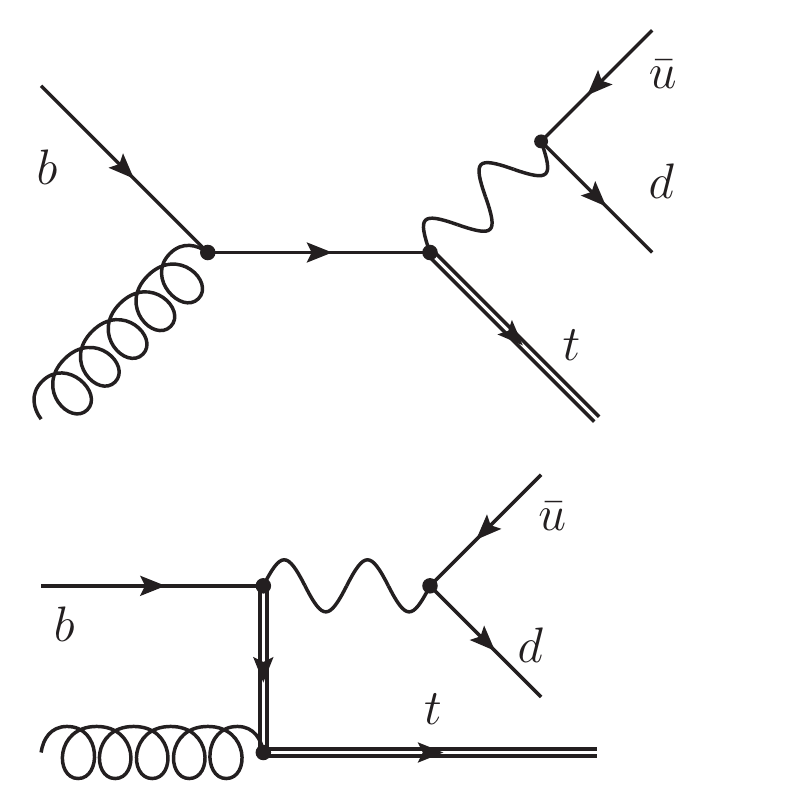}
  \caption{$tW$}
  \label{fig:gq2}
\end{subfigure}
\begin{subfigure}{0.30\textwidth}
  \includegraphics[width=1.0\textwidth]{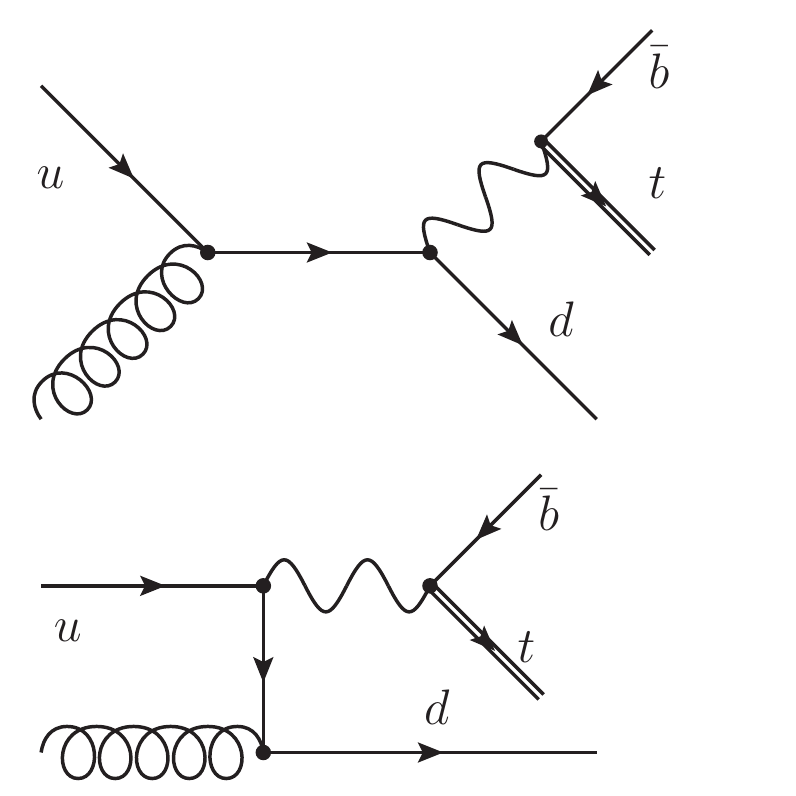}
  \caption{$tjj$}
  \label{fig:gq3}
\end{subfigure}
\caption{Representative LO diagrams classified into $tjj$ and $tW$ production channels.}
\end{center}
\end{figure}

For the remaining subprocesses, a light quark can either come from the
initial gluon (see \fig{fig:gq1}) or from an intermediate $W$ boson
(see \fig{fig:gq2}).  The former diagrams belong to the $tjj$ channel,
while the latter diagrams belong to the $tW$ channel.  As mentioned
before the interference between the two contributions vanishes due to
the different color structures.  Moreover, since each group is
separately gauge invariant, the $tW$-channel diagrams can be
completely removed.  Note that similar diagrams but with
$W \to t\bar{b}$ vertex (see \fig{fig:gq3}) are classified as part of
the $tjj$ channel.
 
Technically, all amplitudes can be obtained via crossing of one
sub-process, say $b u \to t d g$.  For the crossed channel
$g b \to t \bar{u} d$, care must be taken to select only the diagrams
which belong to the $tjj$ channel as explained above.
   
\subsection{Next to leading order QCD}
\label{sect:NLO}
\begin{figure}
\begin{center}
\includegraphics[width=0.8\textwidth]{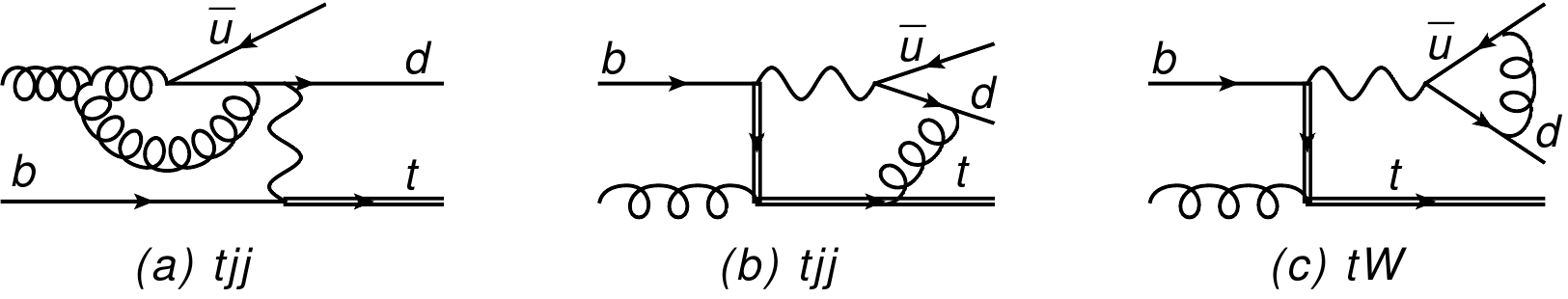}
\caption{Representative one-loop diagrams classified into $tjj$ and $tW$ production channels.}
\label{fig:tjj_virt}
\end{center}
\end{figure}

NLO QCD contributions include virtual and real-emission corrections.
The real-emission processes have one additional parton in the final
state.  Because of this additional QCD emission, color factors of the
$tjj$ and $tW$ amplitudes become more involved and allow interference
between the two contributions. Moreover, the $t\bar{t}$ channel, where
a top quark decays into a $W$ boson and a bottom quark, can lead to the
same final state of $t + 3\text{jets}$.  Furthermore, the interference
between the $tjj$ and the $t\bar{t}$ channels is also non-vanishing.
Similar to what has been done for single top-quark production, these
contributions have to be treated separately to account for the
experimental analysis in which the three processes are analyzed
independently. This can be done in a gauge invariant way by performing
a pole expansion and keeping only non-resonant contributions. Assuming
that experimental cuts will highly suppress these contributions, this
corresponds in practice to dropping resonant diagrams like
Fig.~\ref{fig:tjj_virt} (c). Care must be taken when additional
radiation can lead in general to off-shell contributions and when on-shell
contributions are only generated in certain phase space regions like
for example in Fig.~\ref{fig:tjj_virt} (b). In this case the on-shell
contributions can be extracted using the soft-gluon approximation
\cite{Fadin:1993dz,Melnikov:1995fx,Beenakker:1997ir,Dittmaier:2014qza}
in combination with the complex mass scheme
\cite{Denner:1999gp,Denner:2005fg,Nowakowski:1993iu}. It is well
known that because of soft-collinear factorization real and virtual
corrections cancel each other in the soft limit for sufficiently
inclusive quantities (see for example \bib{Bernreuther:2015fts} where
this is shown for a concrete example). To approximate the non-resonant
part of this contribution, we have combined the respective virtual
corrections with the corresponding real corrections approximated
through the $I$-operator within the Catani-Seymour subtraction
method. Even without applying further experimental cuts to suppress
the $tW$ channel, we find that this contribution gives only a tiny
correction at the sub permille level and it can be safely dropped within
the uncertainties of the final result. For the final results presented in 
\sect{sect:results} and \appen{appen:additional_plots}, this correction is nevertheless included. 

\begin{figure}
\begin{center}
\includegraphics[width=0.95\textwidth]{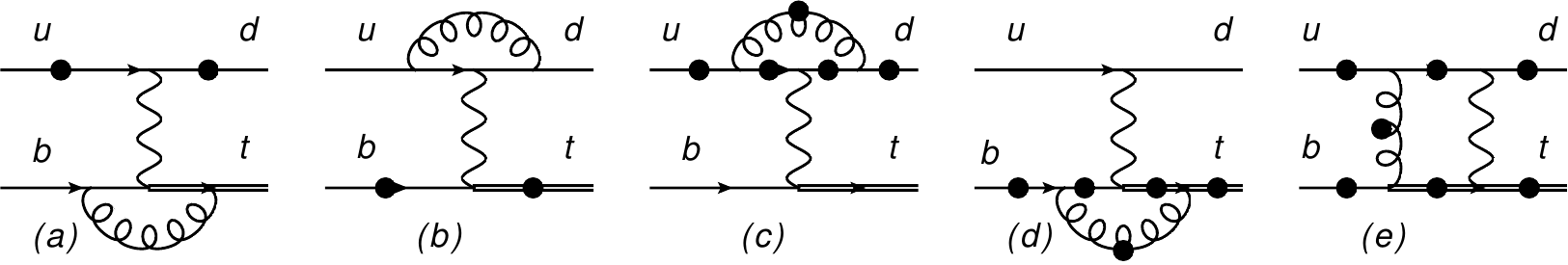}
\caption{Five gauge-invariant groups contributing to the virtual correction of the subprocess $bu\to tdg$. 
The dots represent possible positions for a gluon emission. Only representative diagrams are shown. 
Similar contributions for the other subprocesses 
can be obtained via crossings as for the LO case.}
\label{fig:tjj_virt_5groups}
\end{center}
\end{figure}

We note that all Feynman diagrams for every subprocess can be
classified in gauge-invariant groups. At LO, there are always two
quark lines connected by a $W$ boson exchange. The two gauge-invariant
groups correspond therefore to two cases: either the additional gluon
is connected to the heavy-quark line (i.e. with the top quark) or to
the light-quark line. When a virtual gluon exchange is added, five
gauge-invariant groups arise.  The virtual gluon can be attached
exclusively to the light-quark line as in
Fig.~\ref{fig:tjj_virt_5groups} (b,c) or to the heavy-quark line as in
Fig.~\ref{fig:tjj_virt_5groups} (a,d).  This makes four
gauge-invariant groups. The remaining group corresponds to the case
when there is color exchange between the two quark lines as in
Fig.~\ref{fig:tjj_virt_5groups} (e) (see also Fig.~\ref{fig:tjj_virt}
(b)).  This group contains one-loop five-point integrals and is thus
the most complicated one to calculate. Numerically, this group of
diagrams is strongly suppressed and gives only a tiny contribution to
the full result.

We have performed two independent calculations.
For both calculations, the dipole subtraction method
\cite{Catani:1996vz,Catani:2002hc} is used.  The real-emission
amplitudes are IR divergent in soft and collinear limits. These
singularities have to be regularized and subtracted using subtraction
terms before being integrated over the phase space in four
dimensions. In this step, the $tW$ and $t\bar{t}$ resonant diagrams
are removed, hence interference effects with the $tjj$ channels are
neglected. The subtraction terms are built from the reduced $2 \to 3$
amplitudes keeping only the $tjj$ diagrams as for the LO contribution.
These subtraction terms have to be added back in the form of
integrated dipole contributions called $PK_{tjj}$ and $I_{tjj}$
operators. The IR singularities in these operators are canceled by the
corresponding ones in the PDF counterterms and in the virtual
corrections.

The analytic results have been implemented in two
different computer codes, one written in C++, the other in
Fortran. Extensive cross-checks have been done. We have compared
results at the amplitude level as well as results for the integrated
cross sections and distributions. Within the numerical
uncertainties perfect agreements between the two calculations have
been obtained. Details about these comparisons are provided in
\bib{Moelbitz2019QCD}.

Before presenting the results we would like to provide further details
of the calculation.  In the C++ program, the scalar one-loop integrals
are calculated using the libraries {\tt QCDLoop}~\cite{Ellis:2007qk}
and {\tt FF}~\cite{vanOldenborgh:1990yc}. The $N$-point tensor
integrals are reduced to scalar integrals using the Passarino-Veltman
method~\cite{Passarino:1978jh} for $N\leq 3$ and the tensor-reduction
library {\tt PJFRY}~\cite{Fleischer:2011zz} for the cases $N=4,5$. The
library {\tt PJFRY} uses the methods presented in  \Ref{Fleischer:2010sq}.
This calculation has been cross-checked with a reduction using
\bib{Giele:2004iy} and with {\tt GoSam}~\cite{GOSAM2014}. The
amplitudes for the real corrections  are obtained
using \MG5~\cite{MG5}.  The phase space integration has been done using the
Monte-Carlo integrator {\tt VEGAS}~\cite{Lepage:1978}.

In the Fortran program, scalar one-loop integrals are calculated using
an in-house library {\tt LoopInts} based on the techniques of
\refs{tHooft:1978jhc,Nhung:2009pm,Denner:2010tr}.  $N$-point tensor
integrals are reduced to scalar integrals  using the Passarino-Veltman
method~\cite{Passarino:1978jh} for $N\leq 4$ and 
\bib{Denner:2005nn} for $N=5$. {\tt LoopInts} uses by default double
precision, but will automatically switch to quadruple precision if
numerical instabilities occur in the tensor-reduction.  Helicity
amplitudes are generated using the programs {\tt
  FormCalc}~\cite{Hahn:1998yk}, {\tt FeynArts}~\cite{Hahn:2000kx},
{\tt MadGraph-v4}~\cite{Stelzer:1994ta} which uses {\tt HELAS}
routines~\cite{Murayama:1992gi}.  The $I$ operators required in the
Catani-Seymour subtraction algorithm are implemented
with the help of {\tt AutoDipole}~\cite{Hasegawa:2009tx}.  The
integrator {\tt BASES}~\cite{Kawabata:1995th} is used for the phase-space
integration.

\section{Phenomenological results}
\label{sect:results}

For the input values we use 
\begin{eqnarray}
G_{F} = 1.16638\times 10^{-5} \gev^{-2}, \,
m_W=80.385 \gev, \, m_t = (173.21 \pm 0.87) \gev.
\label{eq:param-setup}
\end{eqnarray}
The masses of all light quarks, {\it i.e.} all but the top-quark mass,
are set to zero.  Partons are combined into jets using the anti-$k_t$
algorithm \cite{Cacciari:2008gp} with the radius $R$ set to $R = 0.4$.
We treat the top-quark as a stable particle and do not include
the top-quark decay. We assume that the top-quark is always tagged and
do not apply the jet-algorithm to the top-quark momentum.  The
momentum of the jet containing the top-quark is thus identified with
the top-quark momentum.  In addition, we impose the following cuts on
the remaining jets: \bea p_{T,j} > 25\gev,\quad |\eta_{j}| < 2.5,
\label{eq:cuts} \eea where $p_{T,j}$ denotes the transverse momentum and
$\eta_{j}$ the pseudo-rapidity of the jet.  For the parton
distribution functions (PDF), we use the {\tt
  PDF4LHC15\_nlo\_100\_pdfas}
set~\bibs{Butterworth:2015oua,Dulat:2015mca,Harland-Lang:2014zoa,
  Ball:2014uwa,Gao:2013bia,Carrazza:2015aoa,Watt:2012tq} via the
library {\tt LHAPDF6}~\bibs{Buckley:2014ana}. For the QCD coupling
constant $\alpha_s$ the value provided by the PDF set, corresponding
to $\alpha_s\left(m_Z\right) = 0.118$ ($m_Z = 91.1876\gev$) for the
chosen PDF, is taken.  The same PDF set is used for both LO and NLO
results.  We produce results for the LHC running at a centre-of-mass
energy of $\sqrt{s}=13\tev$.

\subsection{Inclusive cross sections}
\label{sect:Xsection}
\begin{table}[h]
\renewcommand*{\arraystretch}{1.3}
\centering
\begin{tabular}{|cc|c|c|c|}
  \hline
  $\sigma_{LO}$ & [pb] 
  & $(\delta \sigma)_{\mbox{\scriptsize{PDF}}}$~[\%] & 
  $(\delta \sigma)_{\alpha_s}$~[\%] & $(\delta \sigma)_{m_t}$~[\%]   
  \\ \hline
  $\sigma_{t}$ &22.2 &  
  $\pm 1.7$ & 
  $\pm 1.6$ &
  $\pm 0.8$ 
  \\ 
  $\sigma_{\bar{t}}$ &14.7 &  
  $\pm 2.1$ & 
  $\pm 1.6$ &
  $\pm 0.9$ 
  \\  \hline
  $\sigma_{t+\bar{t}}$ &36.9 &   
  $\pm 1.9$ & 
  $\pm 1.6$ & 
  $\pm 0.8$  
  \\  \hline
\end{tabular}
\caption{
  Inclusive cross sections with PDF, $\alpha_s$ and $m_t$
  uncertainties calculated at LO.
  }
\label{tab:errors_LO}
\end{table}%
Using the aforementioned setup we find in leading order for
$\sqrt{s}=13\tev$ for the production of a single (anti)top-quark in
association with two additional jets the cross sections shown
in \tab{tab:errors_LO}. The cross section for top-quark production is
about 1.5 times larger than the cross section for anti top-quark
production. We also show in \tab{tab:errors_LO} the uncertainties due
to an imperfect knowledge of the PDF's, the QCD coupling constant
$\alpha_s$, and the top-quark mass. 
The PDF and $\alpha_s$ uncertainties are calculated as defined in \bib{Butterworth:2015oua}.
Estimating the PDF uncertainties
using the error PDF's provided by the PDF set, we find an uncertainty
of 1.7 \% for top-quark production and a slightly larger uncertainty
of 2.1\% for anti top-quark production.  For the uncertainties due to
$\alpha_s$ we find in both cases an uncertainty of 1.6\%.  The
uncertainty due to a variation of the top-quark mass within the bounds
allowed by the uncertainty in \eq{eq:param-setup}, leads to an effect of
0.8--0.9\%---consistent with the naive expectation based on the mass
dependence of $t$-channel single top-quark production
\cite{Kant:2014oha}. In conclusion the numerical input is thus
sufficiently well known to allow precise predictions of the cross
section. 

In \tab{tab:Xsections_mt} results for the cross section in
NLO accuracy are given. The quoted values are for
$\mu= \mu_F=\mu_R=m_t$ where $\mu_F$ denotes the factorization scale
and $\mu_R$ the renormalization scale. As central scale the top-quark
mass is used. The NLO corrections enhance the cross section by 22\%
in case of anti top-quark production and almost 30\% in case of
top-quark production. In \tab{tab:Xsections_mt} we have also included
the effects due to a change of the central scale $\mu$ by a factor two
up and down.  The uncertainty is estimated by varying independently
the two scales $\mu_F$ and $\mu_R$ as $n\mu_0/2$ with $n=1,2,4$ and
$\mu_0=m_t$.  The constraint $1/2 \leq\mu_R/\mu_F \leq 2$ is used to
avoid `extrem' scale ratios and associated potentially large
logarithms. This limits the number of possible scale choices to seven
(`seven-point method'). To estimate the uncertainty we determine the
maximal and minimal value for the cross section.  As far as the
leading order cross sections are concerned, this leads to an
uncertainty of about $\pm$10\%. The NLO corrections are thus
significantly larger than the range covered by the leading-order scale
uncertainties, showing that the scale variation does not provide a
reliable uncertainty estimate for the specific cross section. As
expected, the inclusion of the NLO corrections lead to a significant
reduction of the scale uncertainty by roughly a factor two. Compared to
the aforementioned uncertainties related to uncertainties of the PDFs,
$\alpha_s$ and $m_t$, missing higher orders thus provide the dominant
source of uncertainty.
\begin{table}[t]
  \renewcommand*{\arraystretch}{1.3} \centering
  \begin{tabular}{|cccc|cccc|c|}
    \hline
  $\sigma_{LO}$ & [pb] & $(\delta\sigma)_\text{scale}$~[\%] & [pb] & $\sigma_{NLO}$ & [pb] & $(\delta\sigma)_\text{scale}$~[\%] & [pb] & $K$ \\ \hline
  $\sigma_{t}$ & $22.195(2)$ &  $^{+10.5}_{-8.7}$& $^{+2.3}_{-1.9}$ &
  $\sigma_{t}$ & $28.307(8)$ &  $^{+5.8}_{-4.5}$& $^{+1.6}_{-1.3}$ 
  & 1.28
  \\ 
  $\sigma_{\bar{t}}$ & $14.671(1)$ &  $^{+10.5}_{-8.7}$& $^{+1.5}_{-1.3}$ &
  $\sigma_{\bar{t}}$ & $17.856(7)$ &  $^{+4.1}_{-3.8}$& $^{+0.7}_{-0.7}$ 
  & 1.22
  \\  \hline
  $\sigma_{t+\bar{t}}$ & $36.866(2)$ &  $^{+10.5}_{-8.7}$& $^{+3.9}_{-3.2}$ &
  $\sigma_{t+\bar{t}}$ & $46.163(11)$ &  $^{+5.1}_{-4.2}$& $^{+2.4}_{-1.9}$ 
  & 1.25 \\
  \hline
\end{tabular}
\caption{
Inclusive cross sections with scale uncertainties. The central scale is $\mu_0 = m_t$. 
Numbers in the parentheses are statistical errors on the last digits. 
}
\label{tab:Xsections_mt}
\end{table}

\begin{figure}[ht!]
\begin{center}
  \includegraphics[width=0.75\textwidth]{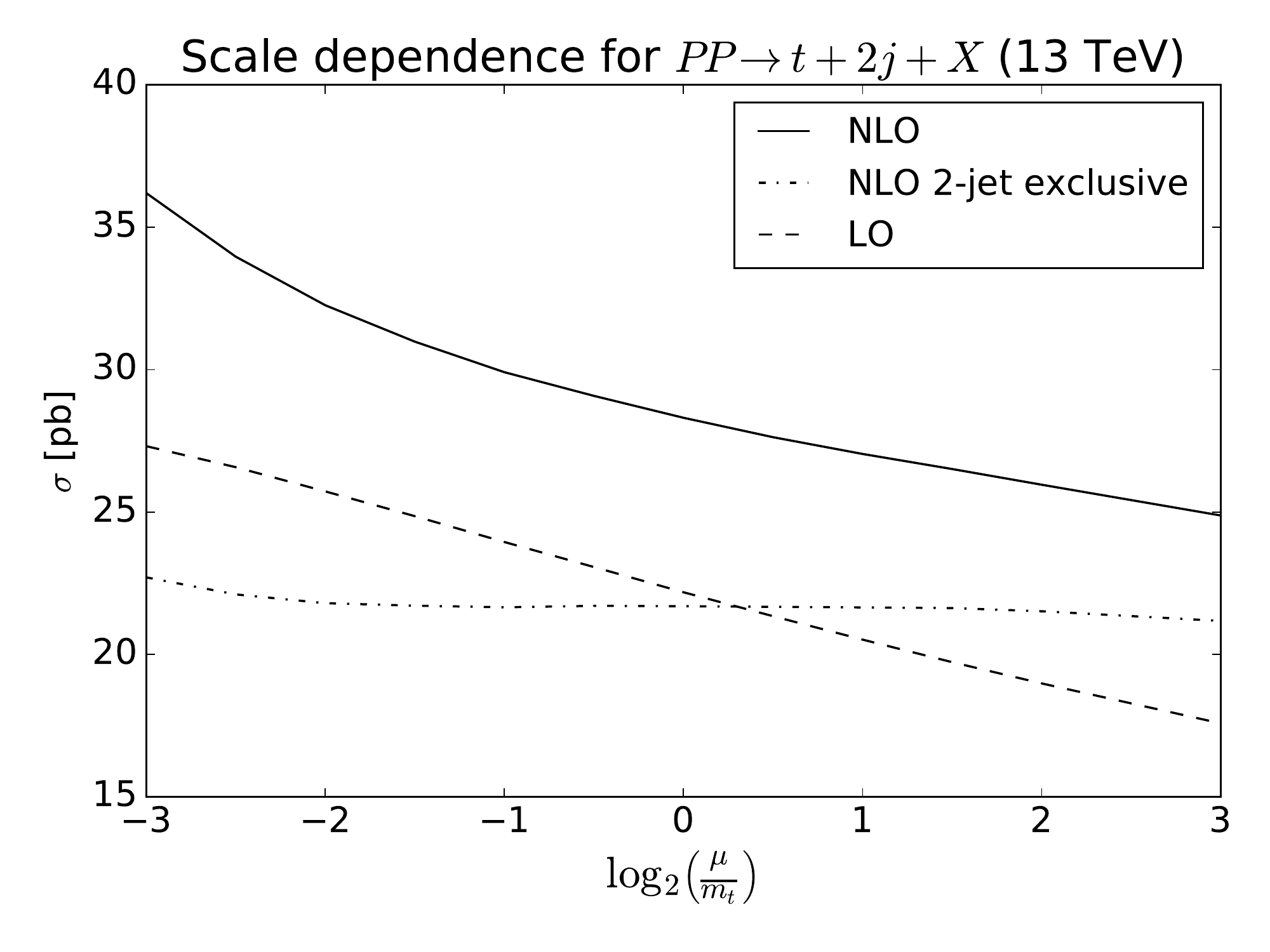}
  \caption{Scale dependence of the inclusive cross sections 
    $\sigma_{LO}$, $\sigma_{NLO}$ and of the two-jet exclusive cross
    section $\sigma_{NLO}^{\mbox{\scriptsize{2jet, exc}}}$.
    Factorization and renormalization scales are set equal, i.e.
    $\mu_F = \mu_R = \mu$.  }
  \label{fig:scaledep_common}
\end{center}
\end{figure}

In \fig{fig:scaledep_common} the scale dependence of the inclusive
cross sections at LO and NLO for single top-quark production in
association with at least two additional jets is shown for the case
$\mu_F = \mu_R = \mu$.  The two-jet exclusive cross section at NLO,
where any additional jet activity is vetoed, is also shown. The
difference between the inclusive and the two-jet exclusive cross
sections gives the contribution of the three-jet events. Note that the
latter is only predicted in LO accuracy.  The corresponding plot for
anti-quark production shows a similar behavior, 
see \fig{fig:scaledep_common-atop} in \appen{appen:additional_plots}. 
As can be seen in
\fig{fig:scaledep_common} even in NLO the cross sections show a scale
dependence typical for a leading-order calculation. In particular, the
scale dependence is not flat.
\begin{figure}[ht!]
\begin{center}
  \includegraphics[width=0.75\textwidth]{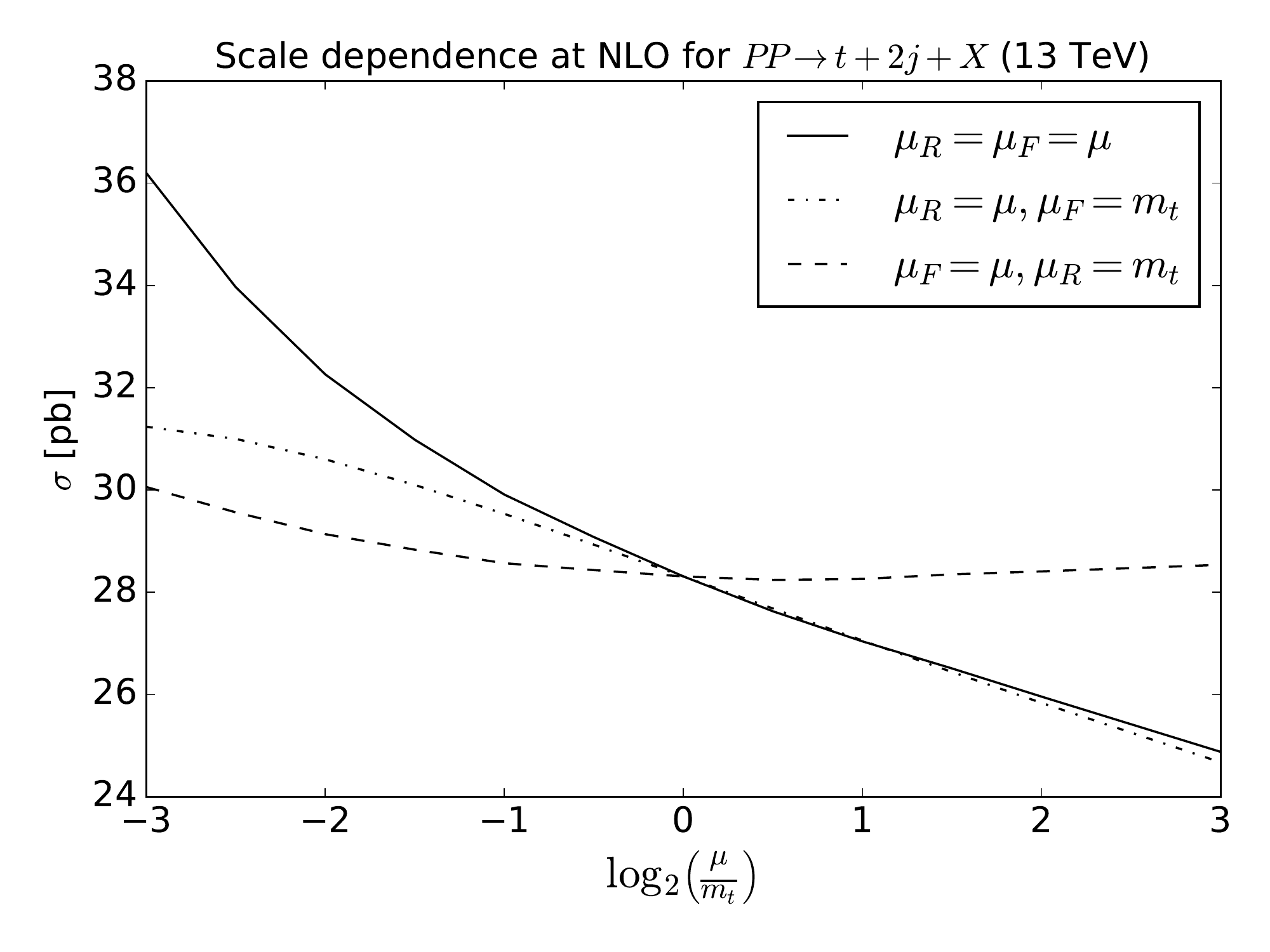}
  \caption{Dependence of the inclusive cross sections $\sigma_{NLO}$ on
    the individual scales $\mu_F$ and $\mu_R$.}
  \label{fig:scaledep_individual}
\end{center}
\end{figure}%
This is because at NLO new channels occur, including in particular
subprocesses with two gluons. These new channels are numerically large
and dominate the scale dependence. To illustrate this effect, the two-jet
cross section where additional jet activity is vetoed is given. The
veto suppresses the contribution from the new channels and leads to a
significantly improved scale dependence of the two-jet exclusive cross
section.  However, this does not necessarily mean that the theoretical
predictions for this observable are more precise.  It is well known
that the jet veto introduces an additional scale and can
lead to additional uncertainties in particular in differential
distributions. This is because the veto scale can lead to large
logarithmic corrections which may spoil the convergence of the
perturbative expansion if not resummed. For more details we refer to
\bib{Stewart:2011cf}. 

In \fig{fig:scaledep_individual} we show results where only one of the
two scales $\mu_F$, $\mu_R$ is changed while the other is kept fixed.
In the range $-1 < \log_2(\mu/m_t) < 3$ the renormalization scale
gives the dominant contribution to the scale dependence. This is
consistent with the aforementioned observation that the scale
dependence is dominated by new channels which occur for the first time
in NLO. Only at rather low scales the factorization scale becomes
important.
\begin{table}[t]
\renewcommand*{\arraystretch}{1.3}
\centering
\begin{tabular}{|c|cc|cc|}
  \hline
  $\sqrt{s}$~[TeV] &  
  $\sigma_{LO}^{t}$ [pb] & $\sigma_{NLO}^{t}$ [pb]  &
  $\sigma_{LO}^{\bar{t}}$ [pb] & $\sigma_{NLO}^{\bar{t}}$ [pb]
  \\ \hline
  $7$  & $7.4^{+0.9}_{-0.8}$ & $8.9^{+0.3}_{-0.3}$ & $4.0^{+0.5}_{-0.5}$ & $4.8^{+0.2}_{-0.2}$\\
  $8$  & $9.6^{+1.1}_{-1.0}$ & $11.6^{+0.4}_{-0.4}$ & $5.4^{+0.7}_{-0.6}$ & $6.5^{+0.3}_{-0.3}$ \\
  $13$ & $22.2^{+2.3}_{-1.9}$ & $28.3^{+1.6}_{-1.3}$ & $14.7^{+1.5}_{-1.3}$ & $17.9^{+0.7}_{-0.7}$ \\
  $14$ & $25.0^{+2.6}_{-2.2}$  & $32.2^{+1.5}_{-1.5}$ & $16.9^{+1.8}_{-1.5}$ & $20.7^{+0.8}_{-0.8}$ \\
  \hline
\end{tabular}
\caption{Cross sections with scale uncertainties at different
  proton-proton colliding energies. }
\label{tab:Xsection_Energ}
\end{table}

In \tab{tab:Xsection_Energ} we show the cross section for different
collider energies. In case of anti--top-quark production the $K$-factor
is only weakly dependent on the collider energy. In contrast, for
top-quark production a significant raise of the $K$-factor can be
observed. As a consequence also the ratio of the two cross sections
depends on the collider energy. At high energies the number of produced
anti--top quarks increases with respect to the number of produced top
quarks. This is because top-quark and anti top-quark production probe
different PDF's with a different energy dependence.

\subsection{Kinematical distributions}
\label{sect:dist}
For the evaluation of the inclusive cross section we
used a fixed renormalization and factorization scale. While this is
appropriate for the total cross section which is dominated by
events with moderate momentum transfer, this is no longer true when
distributions at high momentum transfer are studied. In the latter case
numerically rather different energy scales can occur which may lead to
large logarithmic corrections invalidating the naive use of
perturbation theory. It is well known that in such cases a
dynamical scale often improves the situation.
For the process at hand, we use 
\begin{equation}
  \mu_\text{dyn}= c_d \left( m_t + p_{T,t} + 
    \sum_{i\in\text{partons}} p_{T,i} \right),
 \label{eqn:mudyn}
\end{equation}
as dynamical scale, where $c_d$ is a constant which still needs to be
fixed to a certain value. The above scale choice may be seen as a
modification of $H_T$ often used. The top-quark mass occurring in the
dynamical scale prevents the scale from becoming too small for low
energies, since in this case the top-quark mass provides a
cut-off and should be the relevant energy scale.
\begin{table}
\renewcommand*{\arraystretch}{1.3}
\centering
\begin{tabular}{|lcccccc|c|}
        \hline
  $c_d$ & $2$ & 1 & $1/2$ & $1/4$ & $1/8$ & 
    $1/16$ & $\mu=m_t$\\ \hline
  $\sigma_{LO}^{t}$~[pb] & 18.38 & 19.83 & 21.40 & 23.06 & 24.74 & 26.26& 22.20 \\
  $\sigma_{LO}^{\bar{t}}$~[pb] & 12.12 & 13.12 & 14.20 & 15.34 & 16.49 & 17.56 & 14.67 \\
  \hline
\end{tabular}
\caption{LO cross sections with a dynamical scale using different
  values for the constant $c_d$. For comparison, the result for the
  fixed scale is also given.}
\label{tab:dynScale_LO}
\end{table}
\begin{table}
\renewcommand*{\arraystretch}{1.3}
\centering
\begin{tabular}{|cccc|cccc|c|}
        \hline
  $\sigma_{LO}$ & [pb] & $(\delta\sigma)_\text{scale}$~[\%]  & [pb] & $\sigma_{NLO}$ & [pb] & $(\delta\sigma)_\text{scale}$~[\%] & [pb] & $K$ \\ \hline
  $\sigma_{t}$ & $21.407(2)$ &  $^{+7.7\%}_{-7.3\%}$& $^{+1.7}_{-1.6}$ &
  $\sigma_{t}$ & $27.14(1)$ &  $^{+4.7\%}_{-3.9\%}$& $^{+1.3}_{-1.1}$ 
  & 1.27
  \\ 
  $\sigma_{\bar{t}}$ & $14.197(1)$ &  $^{+8.0\%}_{-7.6\%}$& $^{+1.1}_{-1.1}$ &
  $\sigma_{\bar{t}}$ & $17.23(2)$ &  $^{+3.4\%}_{-3.1\%}$& $^{+0.6}_{-0.5}$ 
  & 1.21
  \\  \hline
  $\sigma_{t+\bar{t}}$ & $35.604(2)$ &  $^{+7.9\%}_{-7.4\%}$& $^{+2.8}_{-2.6}$ &
  $\sigma_{t+\bar{t}}$ & $44.37(2)$ &  $^{+4.2\%}_{-3.6\%}$& $^{+1.9}_{-1.6}$ 
  & 1.25 \\
  \hline
\end{tabular}
\caption{
Inclusive cross sections with scale uncertainties. The central scale is $\mu_\text{dyn}^0$. 
Numbers in the parentheses are statistical errors on the last digit. 
Scale uncertainties are calculated using the three-point $(\mu_\text{dyn}^0/2,\mu_\text{dyn}^0,2\mu_\text{dyn}^0)$ 
method.
}
\label{tab:Xsection_dyn_scale}
\end{table}

To find a reasonable value for the constant $c_d$, we require that the
inclusive cross section matches the one calculated with
$\mu_F = \mu_R = m_t$. In \tab{tab:dynScale_LO} LO results are
provided, varying $c_d$ from $1/16$ to $2$.  The results show that
$c_d=1/2$ is a good choice and reproduces roughly the results obtained
with the fixed scale. We will therefore use hereafter this value to
define the central scale $\mu_\text{dyn}^0$. 

In \tab{tab:Xsection_dyn_scale} we present NLO cross sections together
with the associated scale uncertainties using the dynamical scale with
$c_d=1/2$. The scale uncertainties are estimated using
$\mu_F = \mu_R = \mu_\text{dyn}^0$ and varying the dynamical scale by
a factor 2 up and down.  The results for the inclusive cross section
are in good agreement with the results obtained for a fixed scale
evaluated at $\mu = m_t$. Note that, differently from the scale
uncertainties presented in \tab{tab:Xsections_mt} where the
seven-point method is used, the values in \tab{tab:Xsection_dyn_scale}
are obtained using the three-point method where we identify
$\mu_F=\mu_R$ and the two scales are varied together. This is done to
be consistent with the distributions, where we use the three-point
method to reduce the required computing time.  The uncertainties agree
reasonably well with the ones in \tab{tab:Xsections_mt}. 

\begin{figure}[th!]
  \centering
  \includegraphics[width=0.49\textwidth]{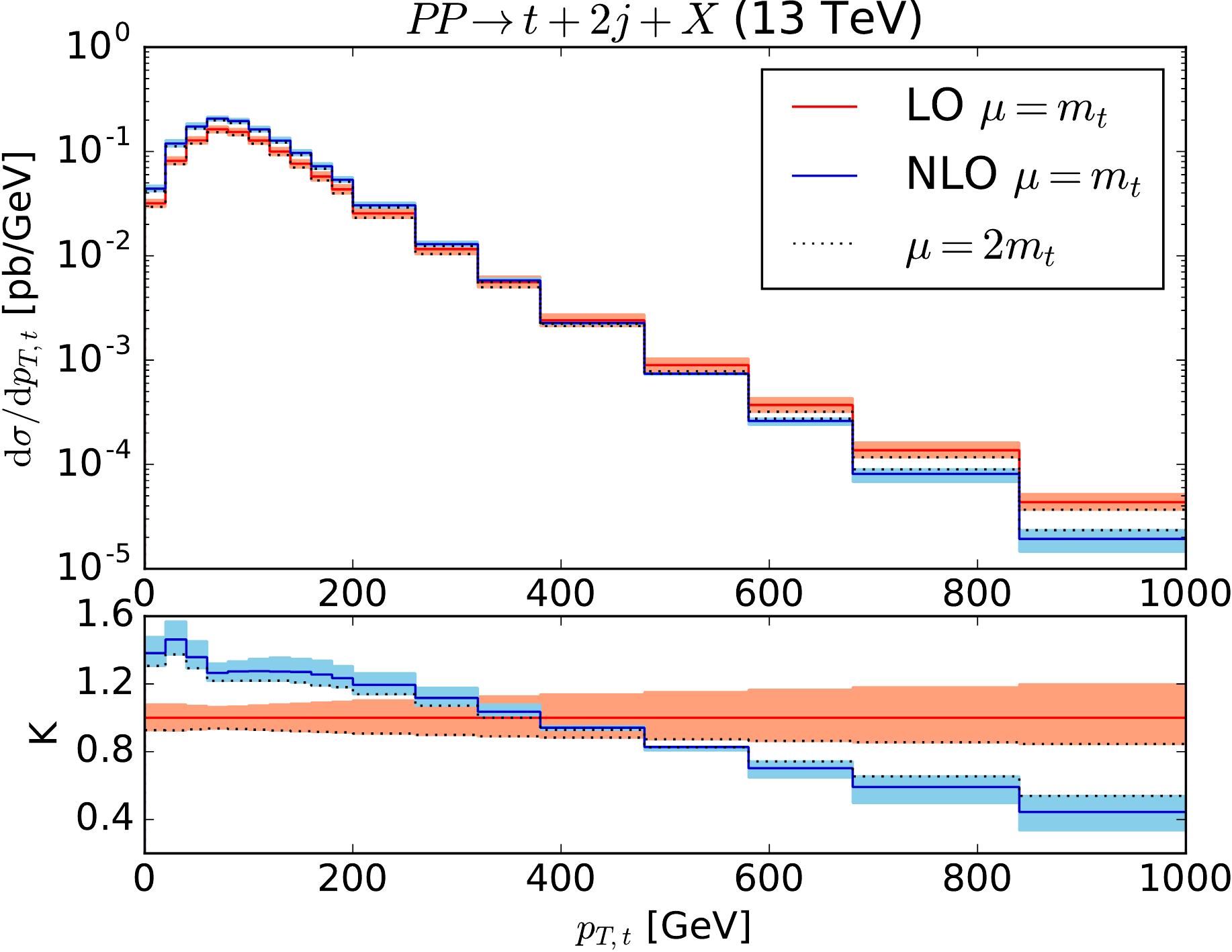}
  \includegraphics[width=0.49\textwidth]{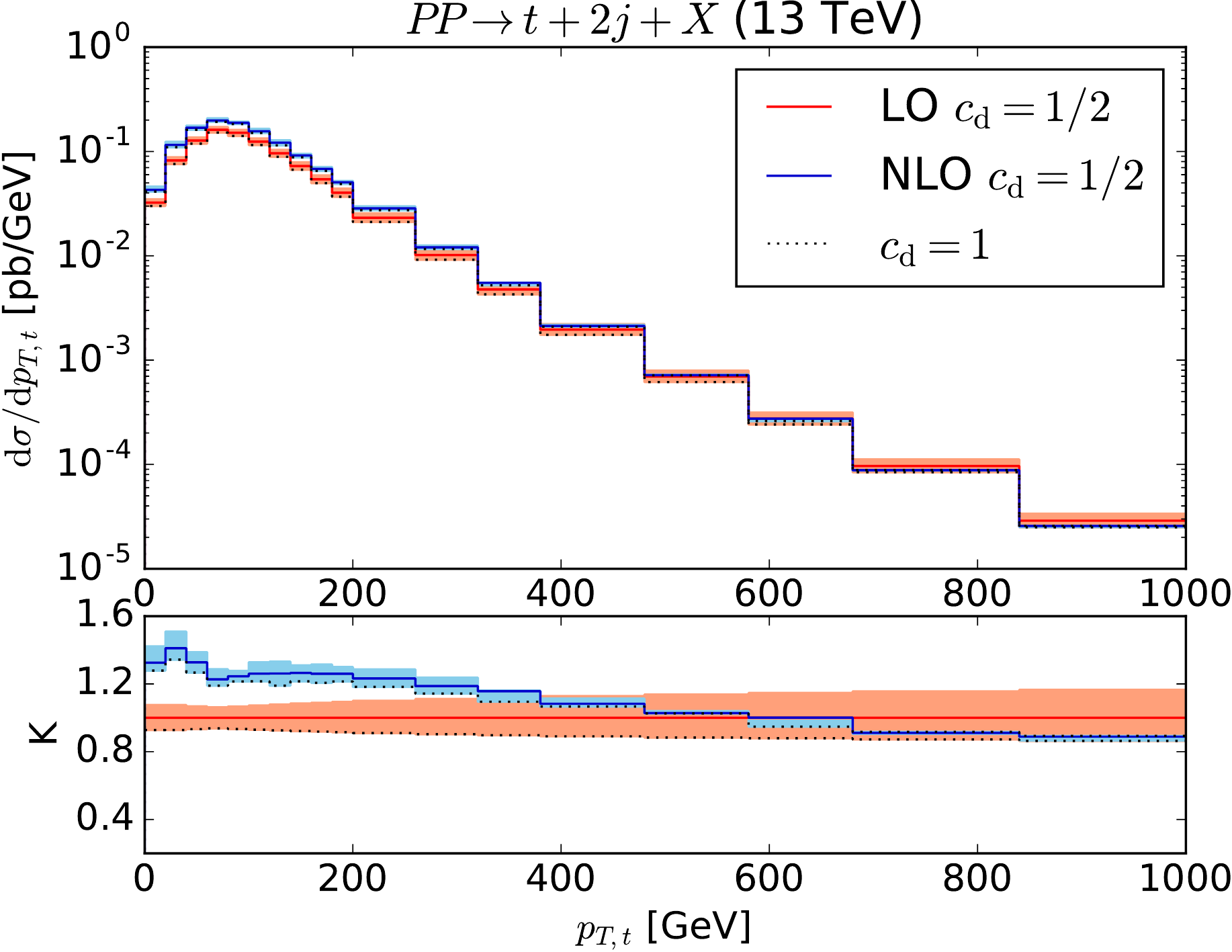}
  \caption{Distributions of the transverse momentum of the top quark with 
    fixed scale $\mu = m_t$ (left) and with dynamical scale $\mu = \mu_\text{dyn}^0$ (right).}
  \label{fig:dist_pT_top_fixed_dyn}
\end{figure}

In \fig{fig:dist_pT_top_fixed_dyn} we compare for the case of the
$p_T$-distribution of the top-quark the two different scale
choices. The left-hand plot shows the result obtained with a fixed
scale as used in the previous section. The right hand plot shows the
same distribution using the dynamical scale.
Using the fixed scale the absolute value of the corrections increases for large
transverse momentum signaling a break down of perturbation theory due
to the appearance of the aforementioned large logarithmic corrections.
This is also partially reflected in the increasing uncertainty
estimated through scale variation. Using the dynamical scale instead,
we find a much improved behaviour. Even at high momentum transfer, the
corrections amount only to $-20$\%. As anticipated, using the
dynamical scale leads thus to a significant improvement of the
perturbative expansion. We stress that the scale choice 
affects only mildly the NLO corrections.
\begin{figure}[th!]
  \centering
  \includegraphics[width=0.6\textwidth]{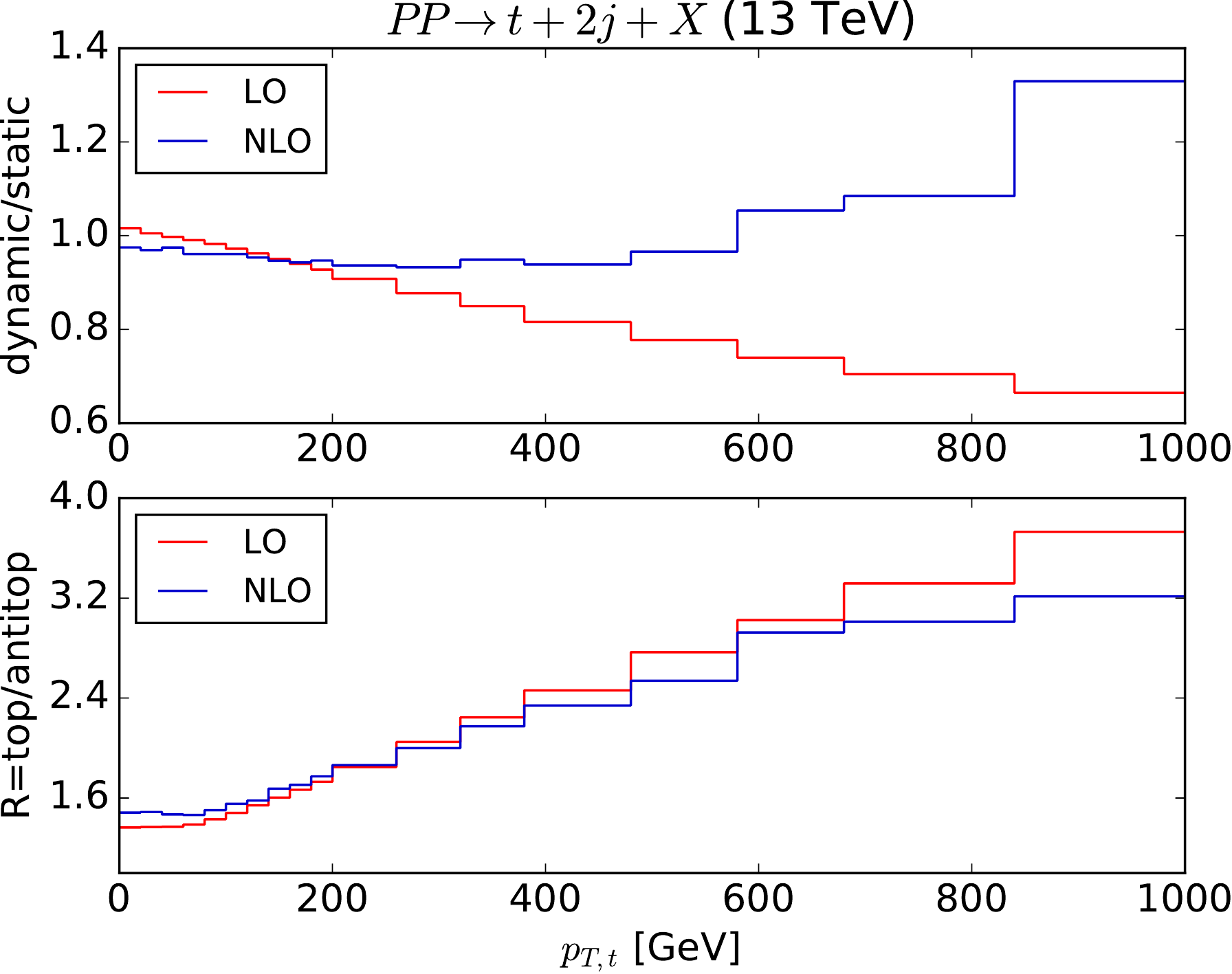}
  \caption{Comparison of predictions using a dynamical scale with
    predictions using a fixed scale (upper plot) for the top-quark production. 
    Ratio between the top-quark and anti--top-quark production (lower plot).}
  \label{fig:dist_pT_top_dyn-vs-stat}
\end{figure}
This is illustrated in the upper plot of
\fig{fig:dist_pT_top_dyn-vs-stat}. Independent of whether a dynamic
scale or a fixed scale is chosen, the NLO corrections are roughly the
same.  The ratio of the two predictions is very close to one. Only
beyond 800 GeV this is no longer true. 
In case the dynamical scale is used, the LO predictions give thus a better
prescription of the full result.

The lower plot of \fig{fig:dist_pT_top_dyn-vs-stat} shows the ratio of the
cross section for top-quark production and the cross section for
anti--top-quark production. The red curve shows the LO result while
the blue curve is the NLO one. The two curves are very close to
each other. However, one can see that the ratio is highly $p_T$
dependent.
At low $p_T$ we recover roughly the factor 1.5 observed for the total
cross section. With increasing $p_T$ the ratio increases. This
information can be used in the experimental analysis to compare for
example top-quark and anti--top-quark tagging efficiencies. 
\begin{figure}[ht!]
  \begin{center}
    \includegraphics[width=0.55\textwidth]{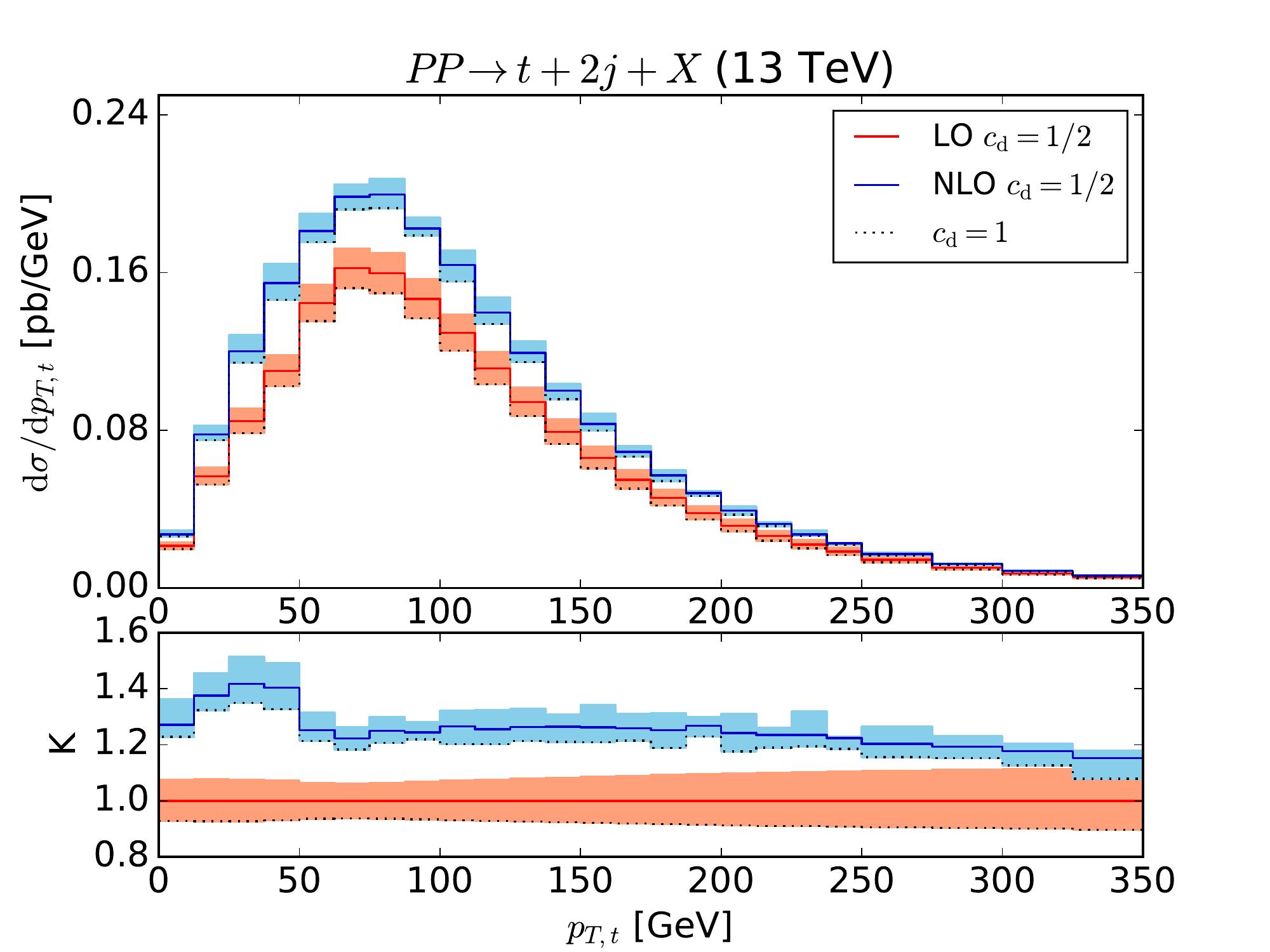}
    \includegraphics[width=0.55\textwidth]{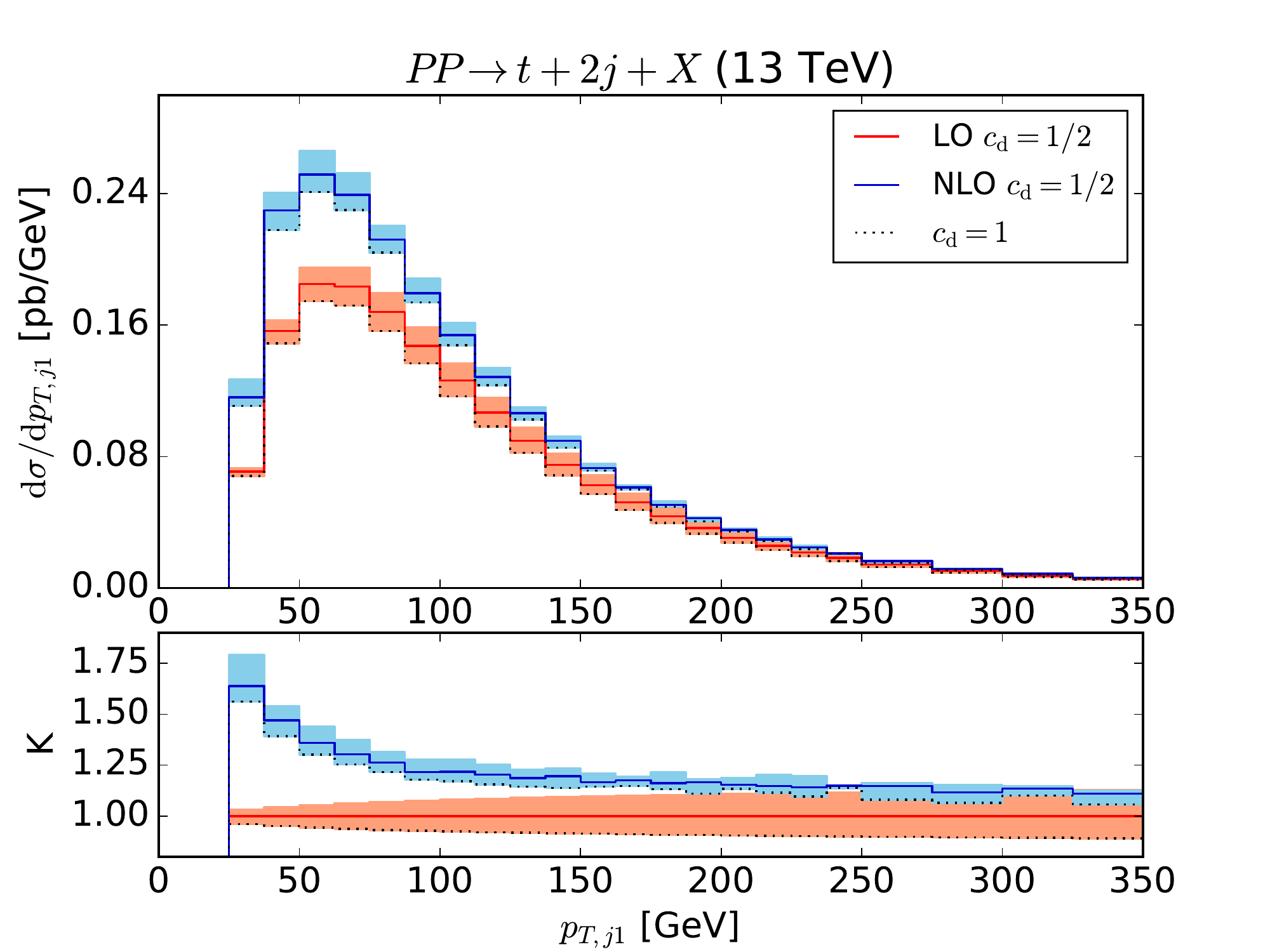} 
    \includegraphics[width=0.55\textwidth]{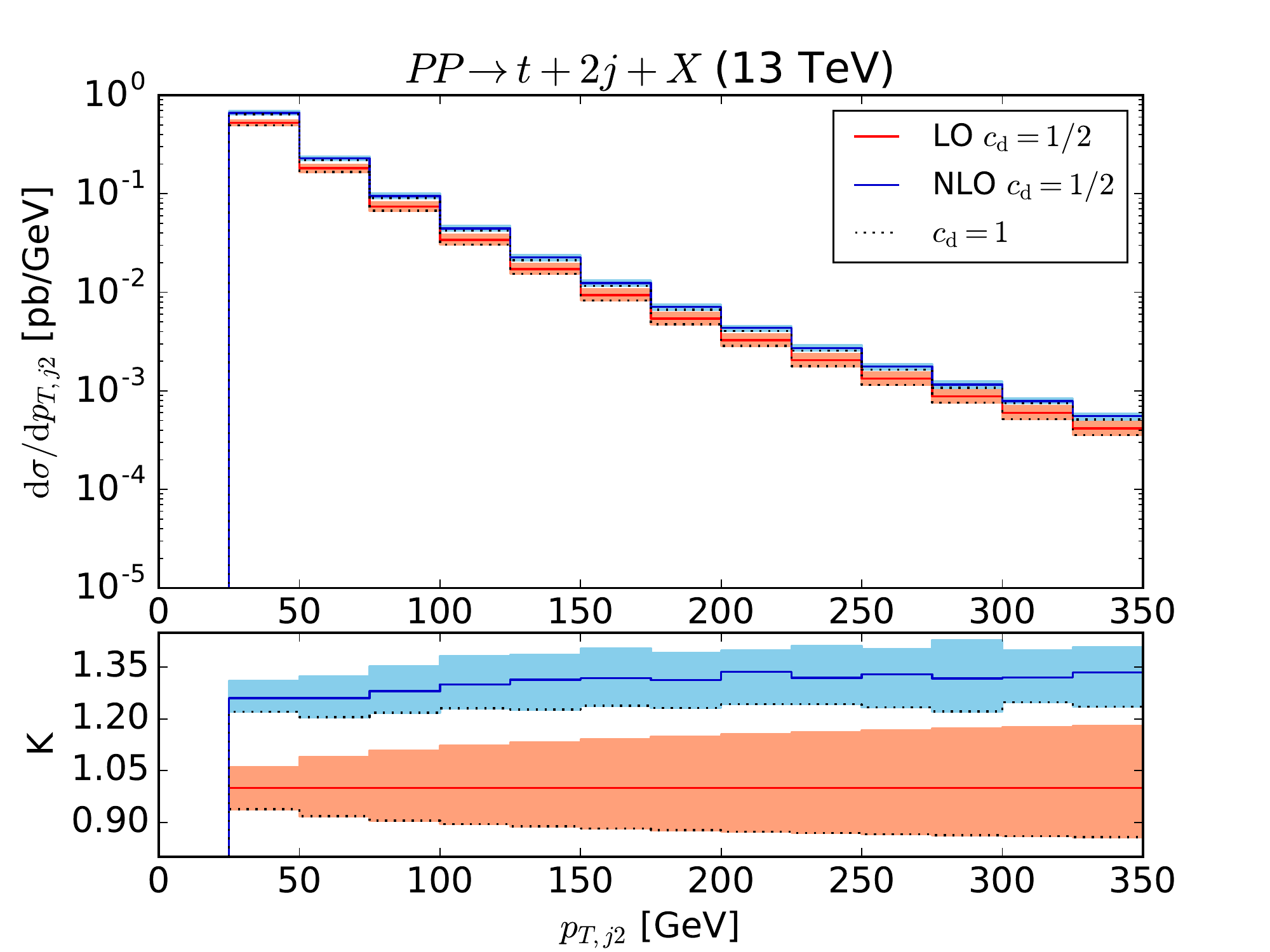} 
  \end{center}
  \caption{Distributions of the transverse momentum of the top quark (top), 
    the first jet (middle), and the second jet (bottom).}
  \label{fig:dist_pT}
\end{figure}

In \fig{fig:dist_pT} we show the transverse momentum distribution for
the top-quark, the first and the second jet. 
The jets are ordered in $p_T$ with the first jet having the largest 
transverse momentum. With exception of the low
momentum region, one observes a flat $K$-factor amounting to positive
corrections of the order 20--30 \%. Most of the jets have a transverse
momentum below 100 \gev. In case of the second jet, the fraction of
jets having a transverse momentum above 100 \gev\ is below 10\%. The
$p_T$-distribution of the top-quark jet peaks at about 75 \gev.  The
$p_T$-distribution of the leading light jet is narrower compared to
the top-quark distribution and peaks at a slightly smaller $p_T$
value. The $p_T$-distribution of the second light jet is a steeply
falling function. The lower end is set by the minimal $p_T$ required
by the jet definition. In case of the $p_T$-distribution of the top
quark the NLO corrections lead to an enhancement below 50\gev. This is
an effect of the real corrections.  The total transverse momentum must
add up to zero. In leading order, the $p_T$ of the top quark must be
compensated by the total $p_T$ of the two additional jets with each
having at least a minimal $p_T$ of 25 \gev\ to pass the cuts. A
top-quark transverse momentum below 50 \gev\ thus restricts the two
additional jets to a very special phase space region. In NLO the total
transverse momentum can be balanced by the third jet. The real
corrections thus lead to an additional positive contribution in the
specific phase space region and the aforementioned enhancement of the
$p_T$-distribution below 50 \gev.  The results for anti top-quark
production are very similar and given in \fig{fig:dist_pT_atop} in \appen{appen:additional_plots}.
\begin{figure}[ht!]
  \centering
  \begin{tabular}{cc}
  \includegraphics[width=0.55\textwidth]{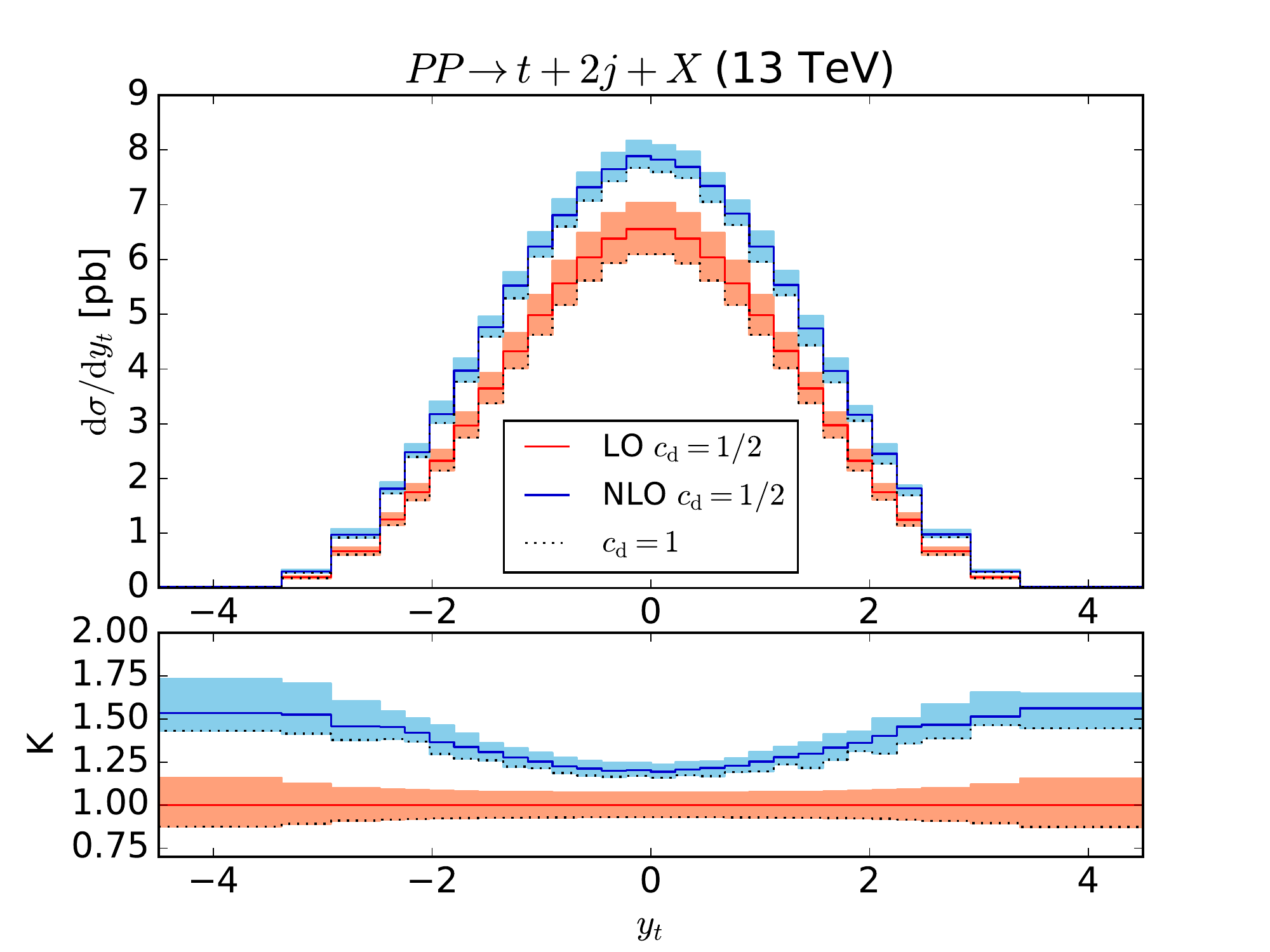}\\
  \includegraphics[width=0.55\textwidth]{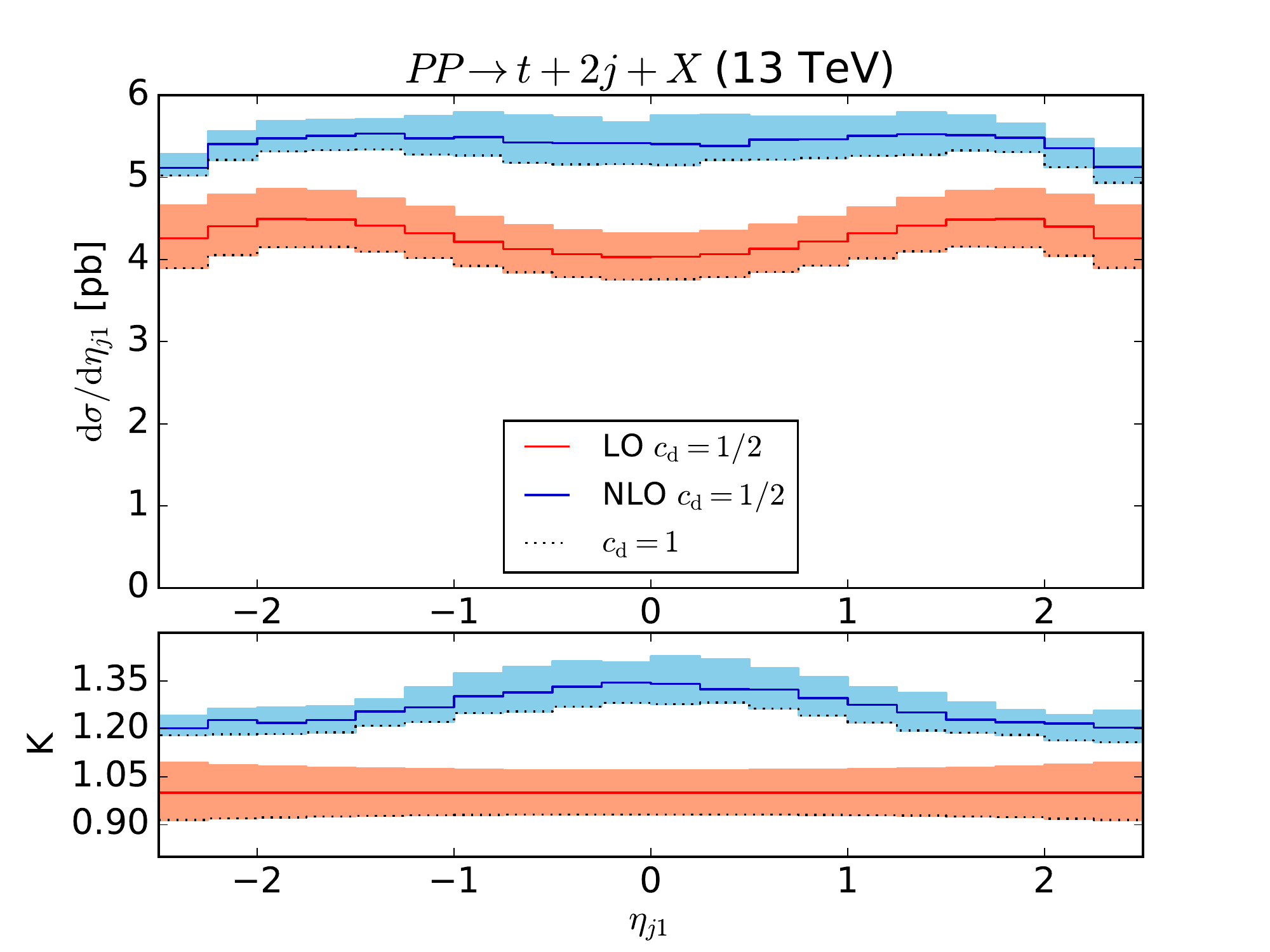}\\ 
  \includegraphics[width=0.55\textwidth]{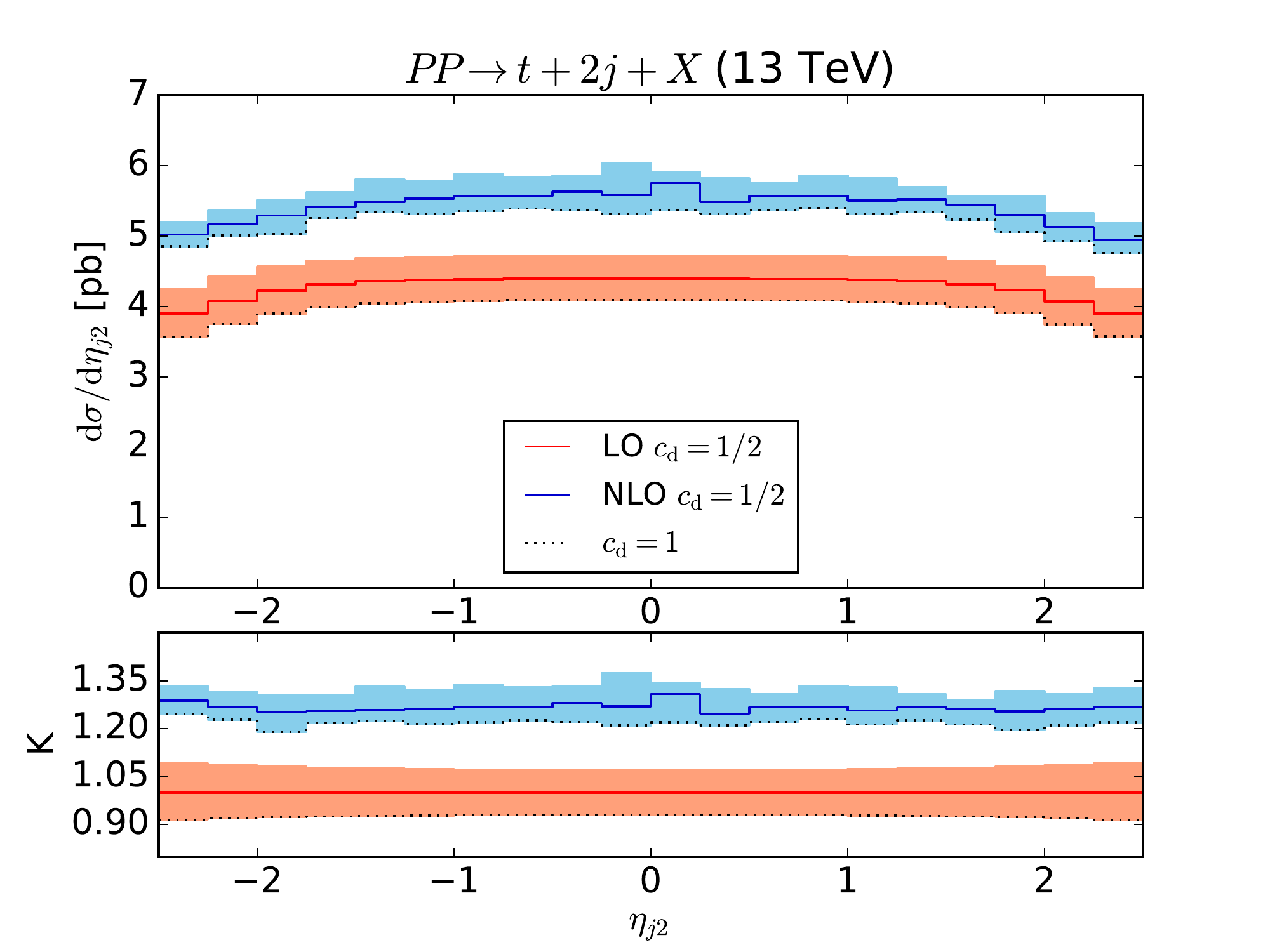}
  \end{tabular}
  \caption{Distributions of the rapidity of the top quark (top), 
    of the pseudorapidity of the first jet (middle) and the second jet
    (bottom).}
  \label{fig:dist_y_eta}
\end{figure}

In \fig{fig:dist_y_eta} we present the rapidity distribution of the
top quark together with the pseudo-rapidity distributions of the two
additional jets. For the top-quark, the corrections for small rapidity
are about 25\% and thus similar to the corrections of the inclusive
cross section and of the $p_{T,t}$-distribution at large transverse
momentum.  For large $|y_t|$ the corrections increase however and are
of the order of 50\%. The origin of this effect is similar to the
effect observed in the $p_T$-distribution of the top quark and is
again a consequence of the real corrections.  A large top-quark
rapidity corresponds to a small value of the top-quark transverse
momentum which requires in leading order again a very special phase
space configuration for the two additional jets.  The additional jet
in the real corrections extends the available phase space and leads
thus to a positive correction to the cross section.  This is also
reflected in the scale dependence.  Since the effect is due to the
real corrections, the results show a large scale dependence.  As far
as the NLO results for the pseudo-rapidity distribution of the two
light jets are concerned, the results look very similar. Both jets
show a rather flat distribution with a slight enhancement for small
$|\eta|$.  In the case of the leading jet, a dip occurring for
$\eta_{j1}=0$ in leading-order is mostly washed out by the NLO
corrections. At NLO only a minor depletion is visible for small
$|\eta|$. The origin of this effect has been traced back to the
three-jet event contribution of the quark-gluon induced channels
(quark $\ne b$), with the $ug$ subprocess being the dominant
contribution.  This is consistent with the observation that the NLO
scale uncertainties are large around $\eta_{j1} = 0$.  Similar results
for the anti top-quark production are shown in
\fig{fig:dist_y_eta_atop} in in \appen{appen:additional_plots}. We
observe that there the dip in the LO $\eta_{j1}$ distribution is much
less pronounced compared to the top-quark production.  This is most
probably due to the difference in the $u$ and $d$ PDFs.
\begin{figure}[ht!]
  \centering
  \includegraphics[width=0.6\textwidth]{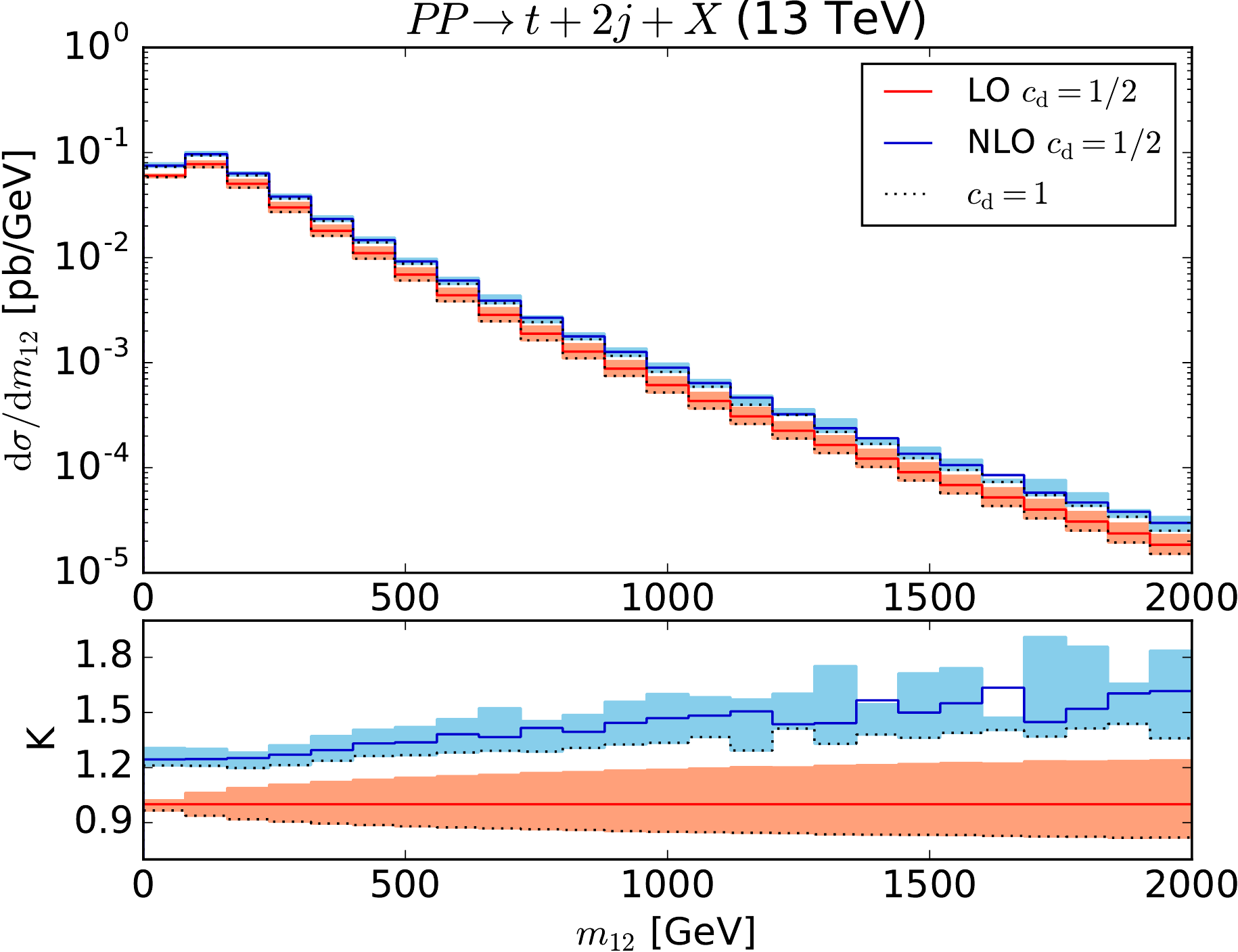}
  \caption{Distribution of the invariant mass
  $m_{12}=\sqrt{(p_{j1}+p_{j2})^2}$
  of the two leading jets. }
  \label{fig:m_jj}
\end{figure}

In \fig{fig:m_jj} the distribution of the invariant mass of the two
light jets is shown. For small $m_{12}=\sqrt{(p_{j1}+p_{j2})^2}$ one
observes again positive corrections of about 20\%. The corrections
increase with increasing $m_{12}$. At the same time the scale
uncertainty---although reduced compared to the leading order
results---becomes larger. The increasing corrections are 
due to the real corrections. We have checked that the virtual corrections ($I$ operator 
included) are negative and are not responsible for that behaviour. 
Similar results for the anti top-quark production are provided 
in \fig{fig:m_jj_atop} in \appen{appen:additional_plots}.

\begin{figure}[ht!]
  \centering
  \begin{tabular}{cc}
  \includegraphics[width=0.6\textwidth]{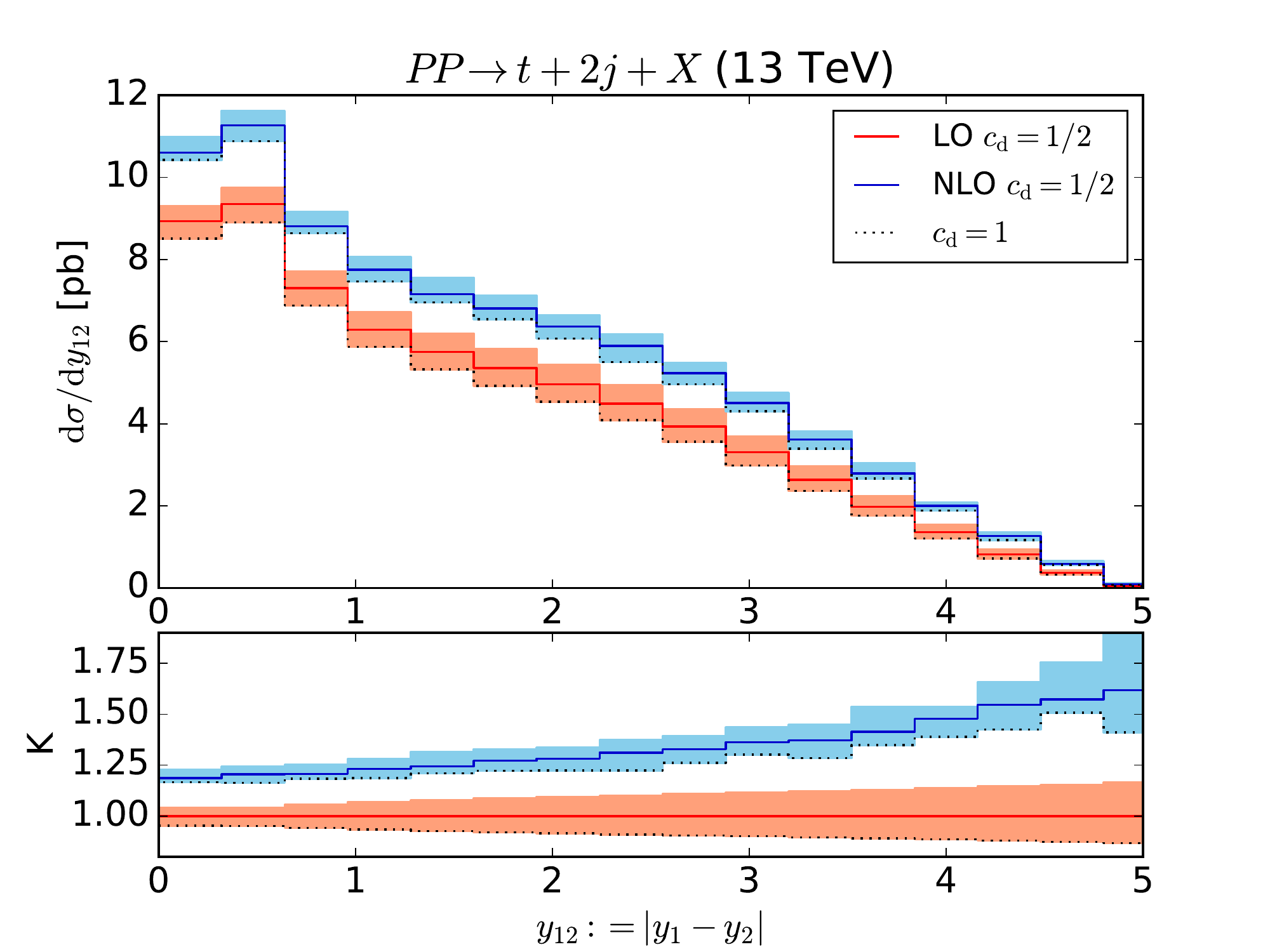}\\
  \includegraphics[width=0.6\textwidth]{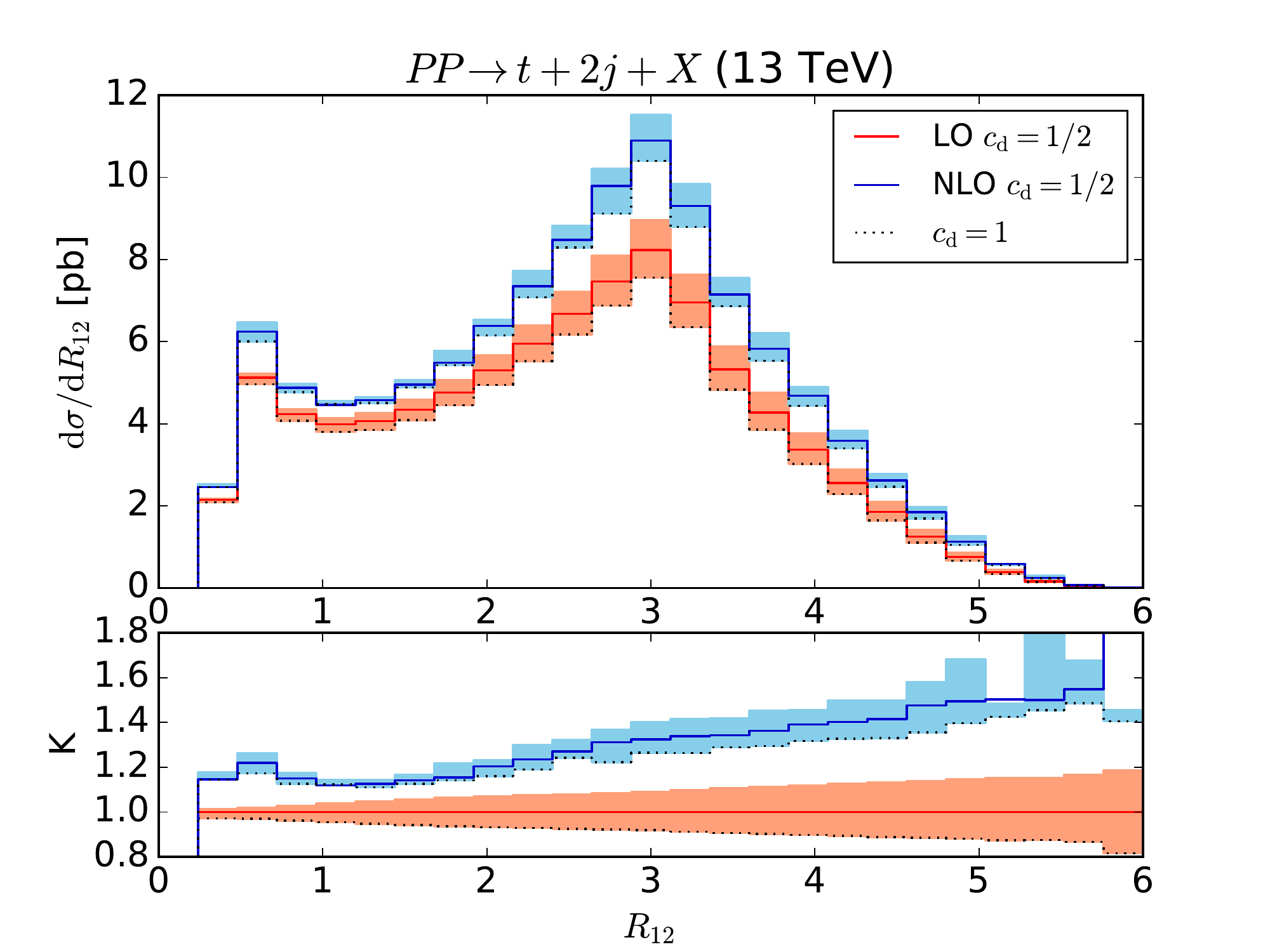}\\
  \end{tabular}
  \caption{Distributions 
    of the rapidity separation (top) and  the $R$ separation (bottom) 
    of the two leading jets. 
  \label{fig:dist_jj}}
\end{figure}

To quantify the spacial separation between the two leading
jets we define 
\begin{equation}
  \Delta y_{12} = |y_1 - y_2|, 
\label{eq:Delta_y}
\end{equation}
and
\bea
\Delta R_{12} = \sqrt{ (y_1 - y_2)^2 + (\phi_1 - \phi_2)^2}. 
\label{eq:Delta_R}
\eea
In \fig{fig:dist_jj} the distributions of $y_{12}$ and $R_{12}$ 
are shown.
As one can see from the upper plot, a small rapidity difference
between the two leading jets is the preferred configuration. The lower
plot of \fig{fig:dist_jj} shows the distributions of the distance in
the $y-\phi$ plane. The LO and NLO $\Delta R_{12}$ distributions peak
around $\pi$, which corresponds, for small rapidity differences, to
the configuration that the two jets are back-to-back. There is a
second less pronounced peak around 0.5. In this case the two jets
recoil against the top quark.  Note that, $\Delta R_{12}$ must be
larger than 0.4 because of the jet definition.  For moderate
$\Delta y_{12}$ and $\Delta R_{12}$ the NLO corrections to the
$\Delta y_{12}$ and $\Delta R_{12}$ distributions tend to be slightly
larger than for the inclusive cross section. The corrections increase
for large $\Delta y_{12}$ and large $\Delta R_{12}$. As can be seen in
\fig{fig:dist_jj} the scale uncertainties increase together with the
size of the corrections. Again this is an effect of the real
corrections. The corresponding distributions for the anti top-quark
production show a similar behavior and is given in
\fig{fig:dist_jj_atop} in \appen{appen:additional_plots}.
 
\clearpage
\newpage

\section{Conclusions}
\label{sect:conclusion}
We have presented a calculation of the NLO QCD corrections for
single on-shell top-quark production in association with two jets at
the LHC.  It is assumed that the $tW$ production mode is measured
separately.  At LO, the interference between the $tjj$ and the $tW$
channels vanishes because of different color structures.  At NLO,
additional QCD radiation introduces interference effects between $tjj$
and $tW$, and also with the $t\bar{t}$ production with one top quark
decaying into three jets. However, these production modes peak in
different phase-space regions, hence interference effects are expected
to be very small, in particular when experimental cuts to separate the
different channels are applied. We have checked that within the soft-gluon
approximation the contribution is indeed tiny. With this assumption, the $tjj$
contribution can be measured independently.

Using inclusive cuts of $p_{T,j} > 25\gev$, $|\eta_{j}|<2.5$ and the
anti-$k_t$ algorithm with a radius $R=0.4$ to define jets, the NLO QCD
corrections for the cross section at 13 TeV are about 28 (22)\% for
top (anti-top) quark production.  The theoretical uncertainties are
dominated by missing higher order contributions, which are estimated,
using a variation of the renormalization and factorization scales, to
be about 5\% at NLO. Uncertainties due to an imperfect knowledge of
the PDFs and of the strong coupling constant are about 2\% at LO.

Further predictions for various kinematical distributions have been
provided.  Using a well-motivated dynamical scale choice for the
renormalization and factorization scales, in most cases moderate $K$-factors are
observed, showing similar corrections as the inclusive
cross section.  However, the QCD corrections have a non-trivial
dependence on the phase-space leading to large corrections in specific
phase-space regions.  For example, for the $p_T$ distribution of the
top quark, the correction is about $+40\%$ in the region of
$p_T \le 50\gev$, then drops to about $+20\%$ for $50 < p_T < 300\gev$
before decreasing steadily with high energies.  Corresponding results
for the anti top-quark production have also been presented, thereby
allowing for comparisons between the two production modes.

The results presented here provide one of the missing building
blocks towards the next-to-next-to-leading order QCD corrections for
single top-quark production beyond the leading color approximation.

\acknowledgments

The work has been partly
supported by the German Ministry of Education and Research under contract no.
05H15KHCAA. The work of LDN is funded by the Vietnam
National Foundation for Science and Technology Development (NAFOSTED)
under grant number 103.01-2017.78.

\appendix
\section{Results for anti top-quark production}
\label{appen:additional_plots}
\begin{figure}[htbp]
  \begin{center}
    \includegraphics[width=0.75\textwidth]{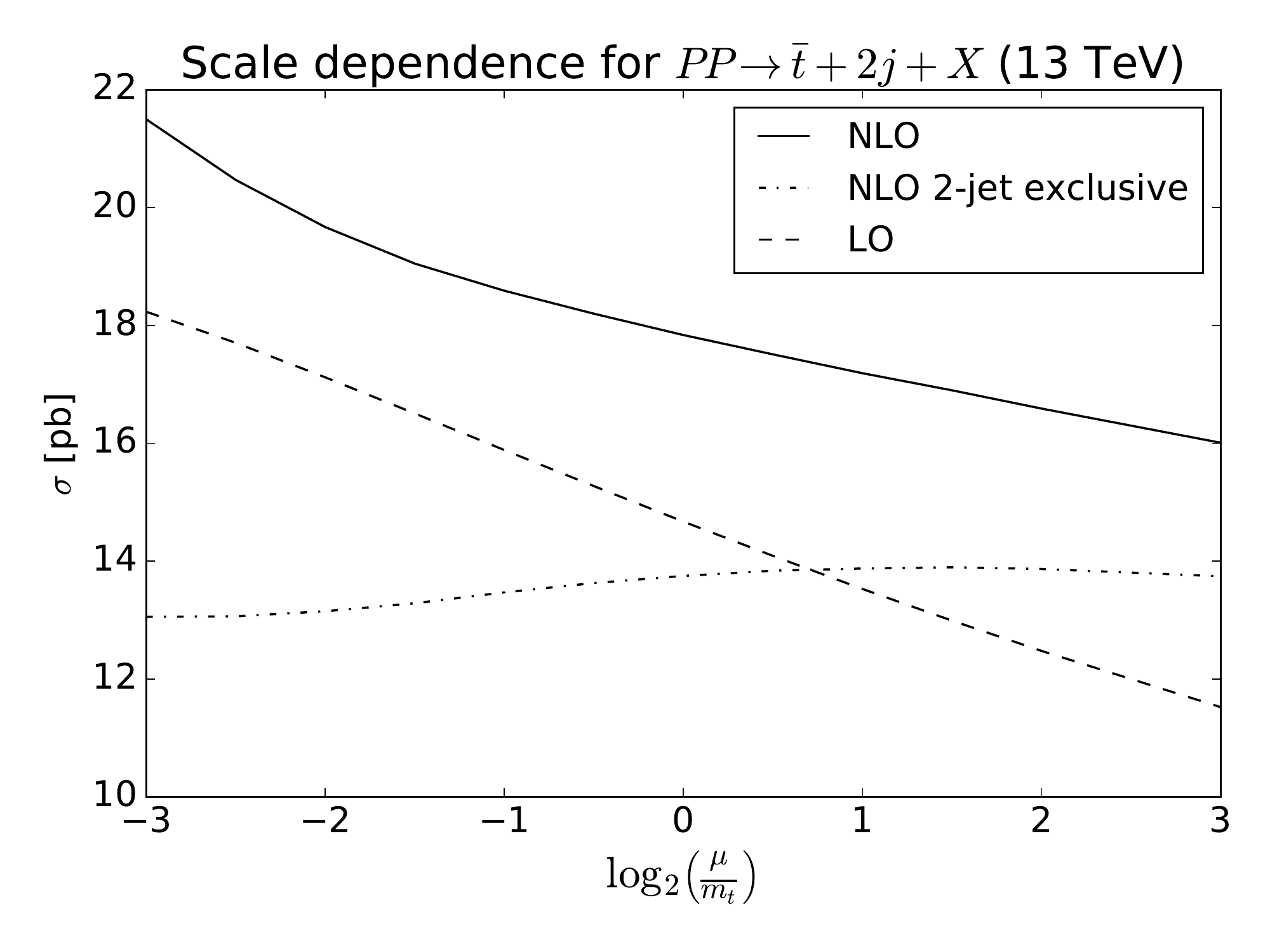}      
  \end{center}
  \caption{Same as \fig{fig:scaledep_common} but for anti top-quark
    production.
    \label{fig:scaledep_common-atop}}
\end{figure}

\begin{figure}[htbp]
  \begin{center}
    \includegraphics[width=0.55\textwidth]{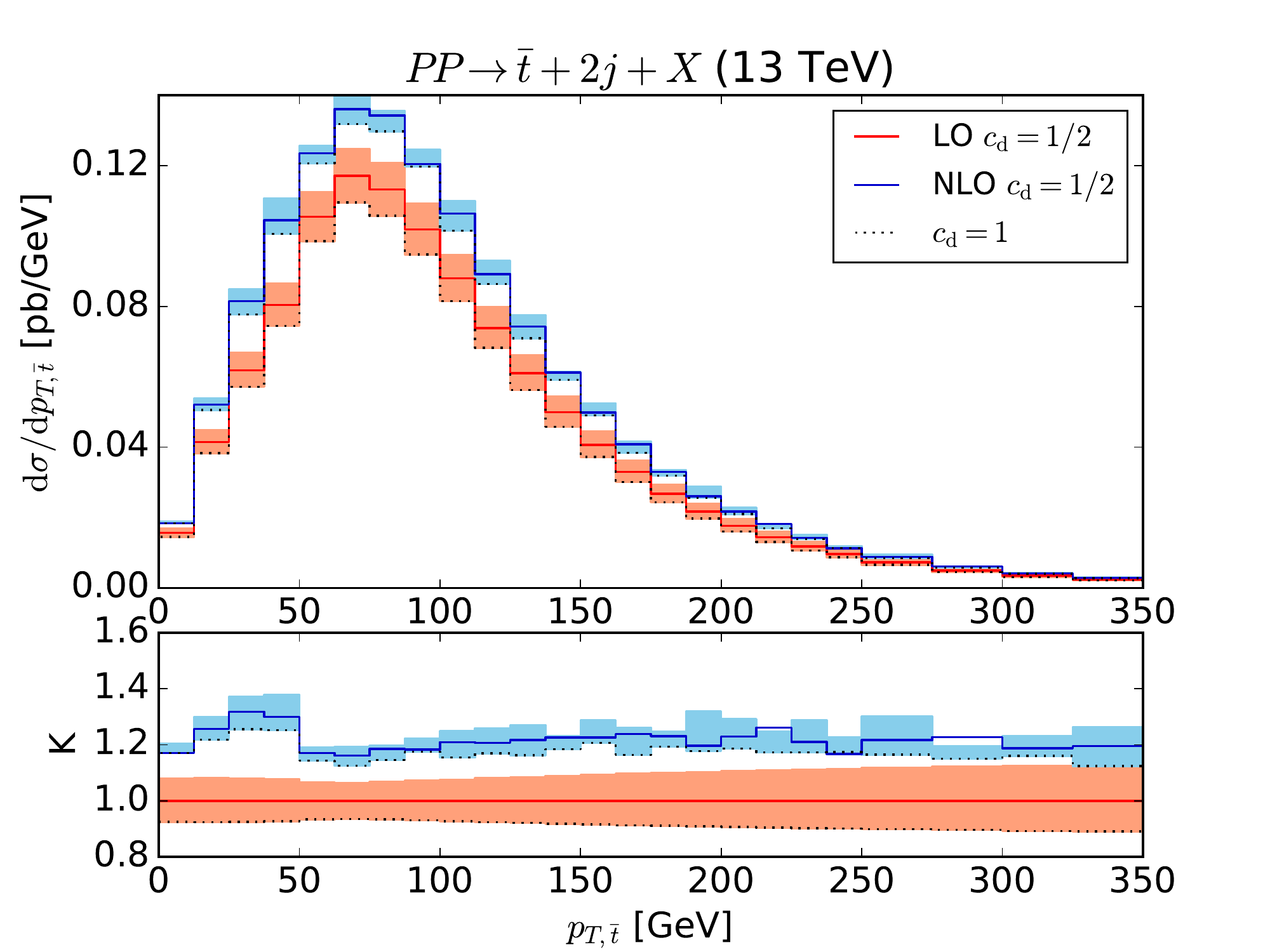}
    \includegraphics[width=0.55\textwidth]{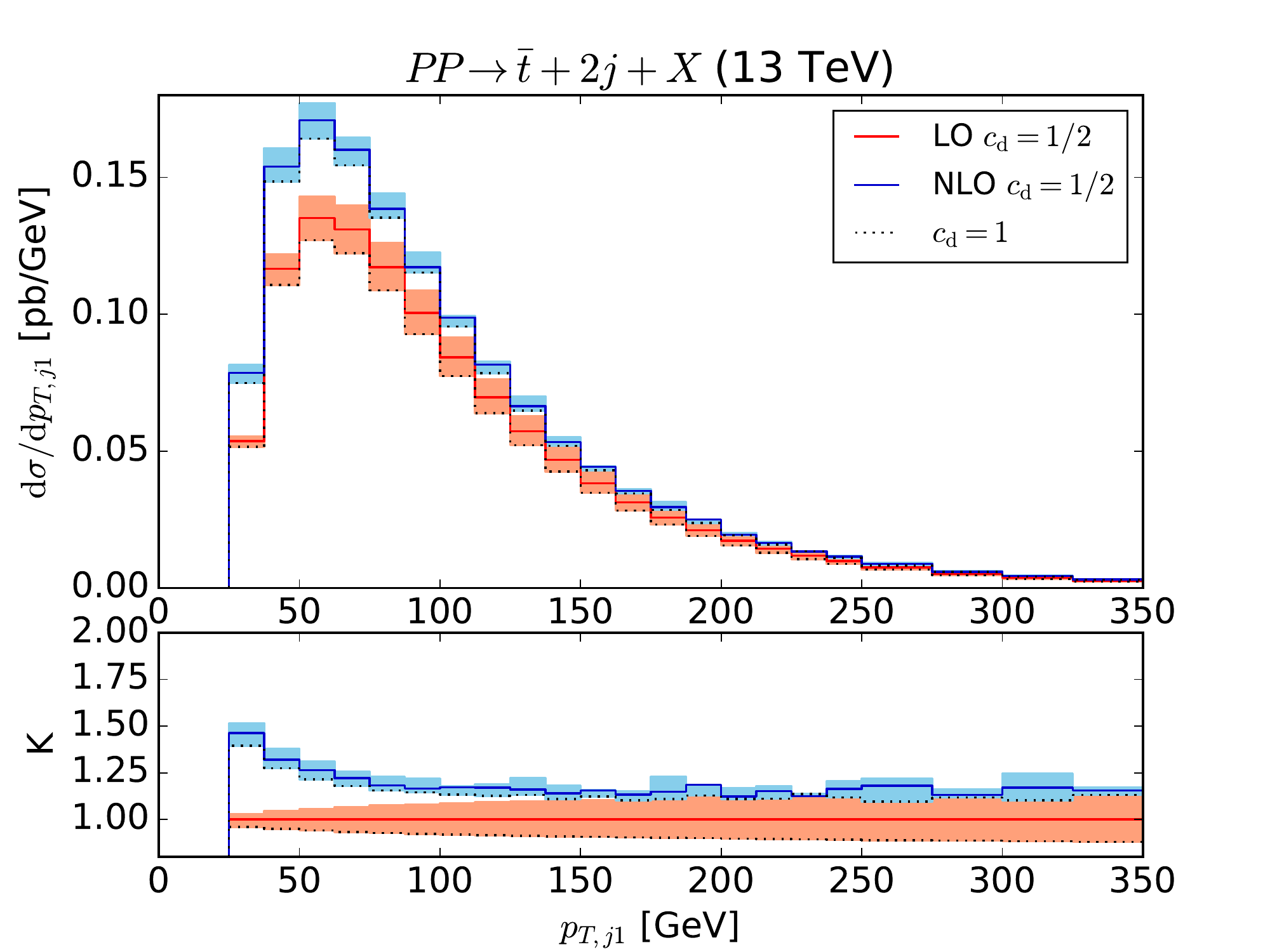}
    \includegraphics[width=0.55\textwidth]{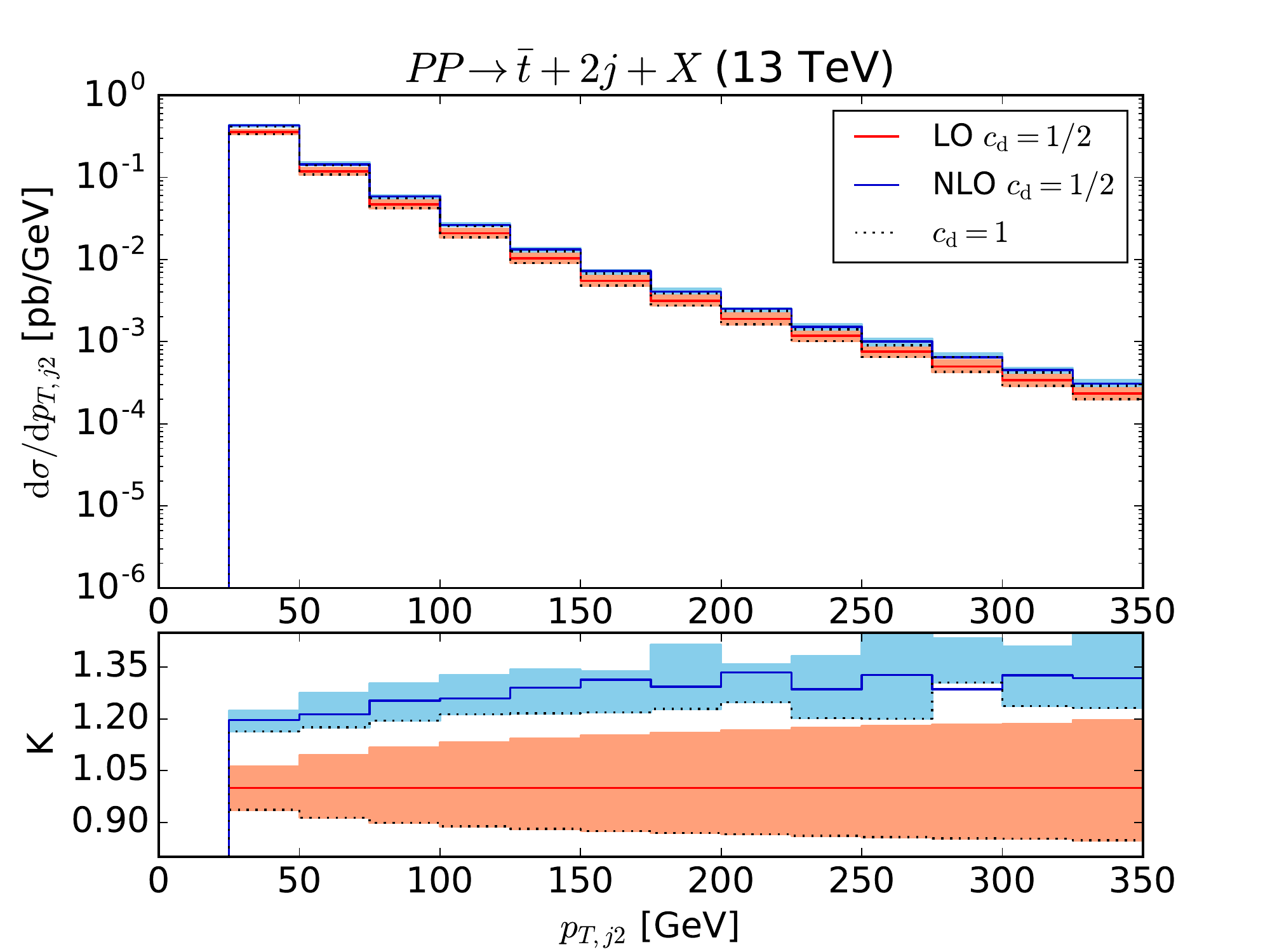}    
  \end{center}
  \caption{Same as \fig{fig:dist_pT} but for anti top-quark
    production. \label{fig:dist_pT_atop}}
\end{figure}

\begin{figure}[ht!]
  \centering
  \begin{tabular}{cc}
  \includegraphics[width=0.55\textwidth]{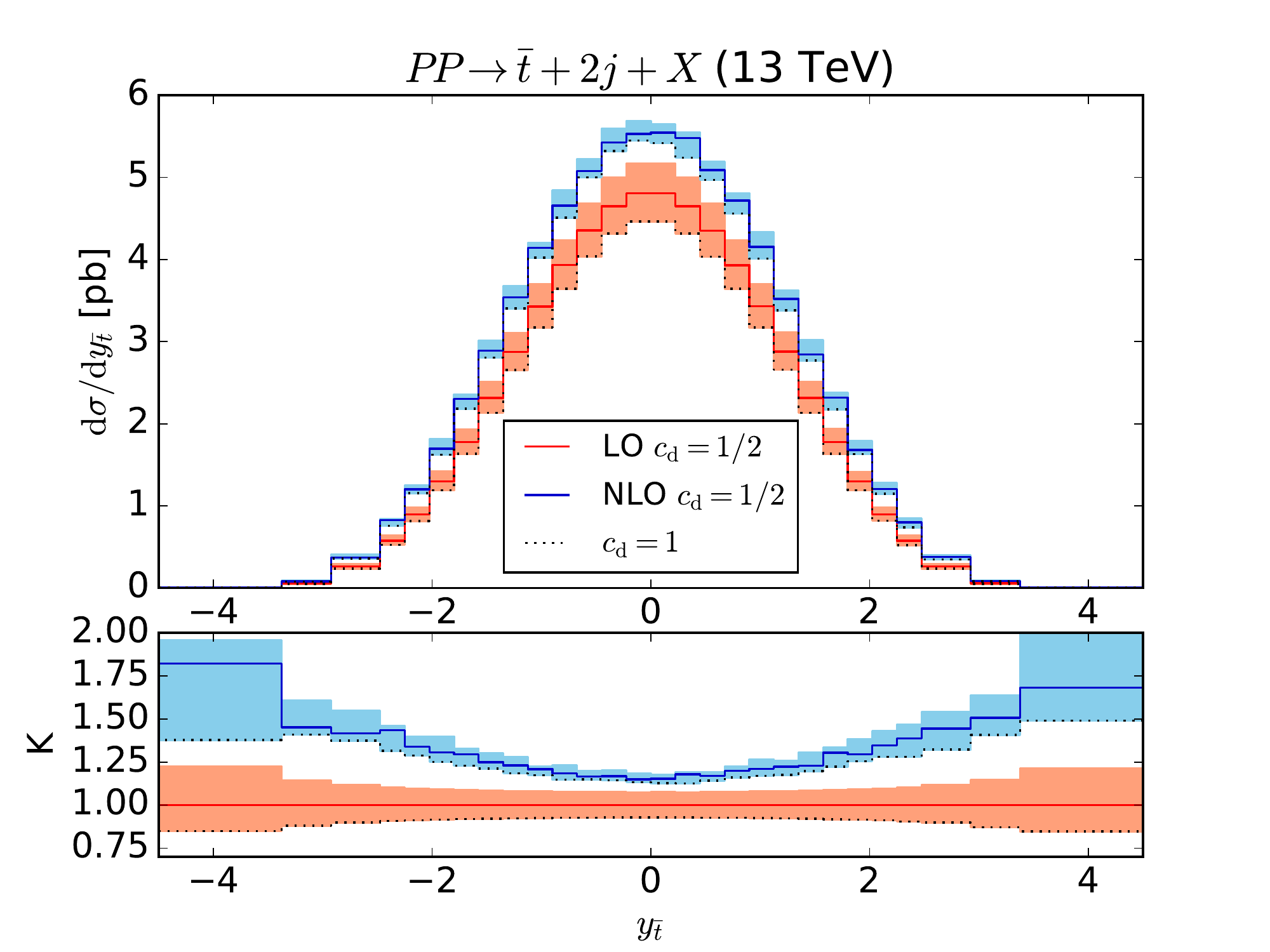}\\
  \includegraphics[width=0.55\textwidth]{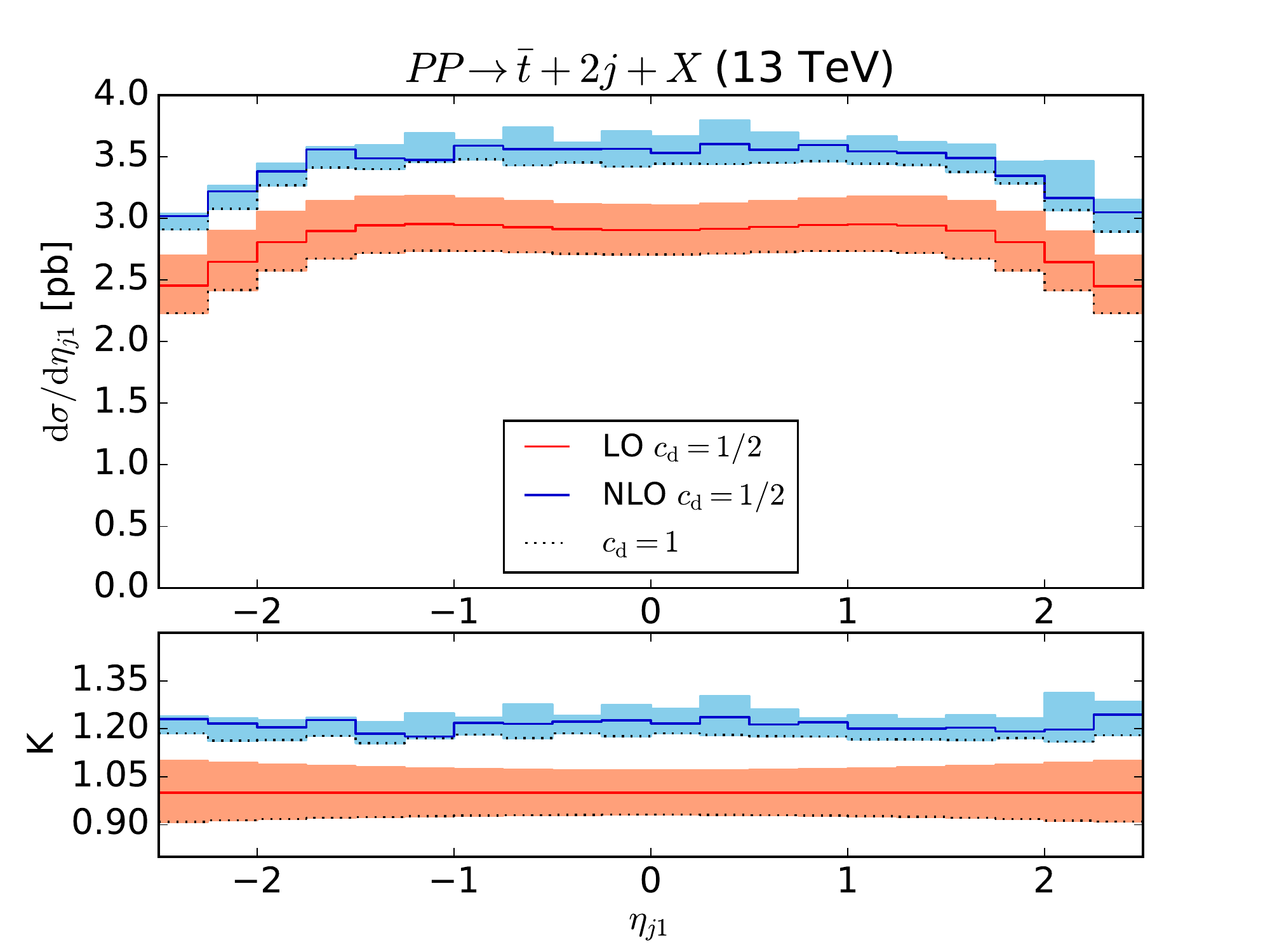}\\ 
  \includegraphics[width=0.55\textwidth]{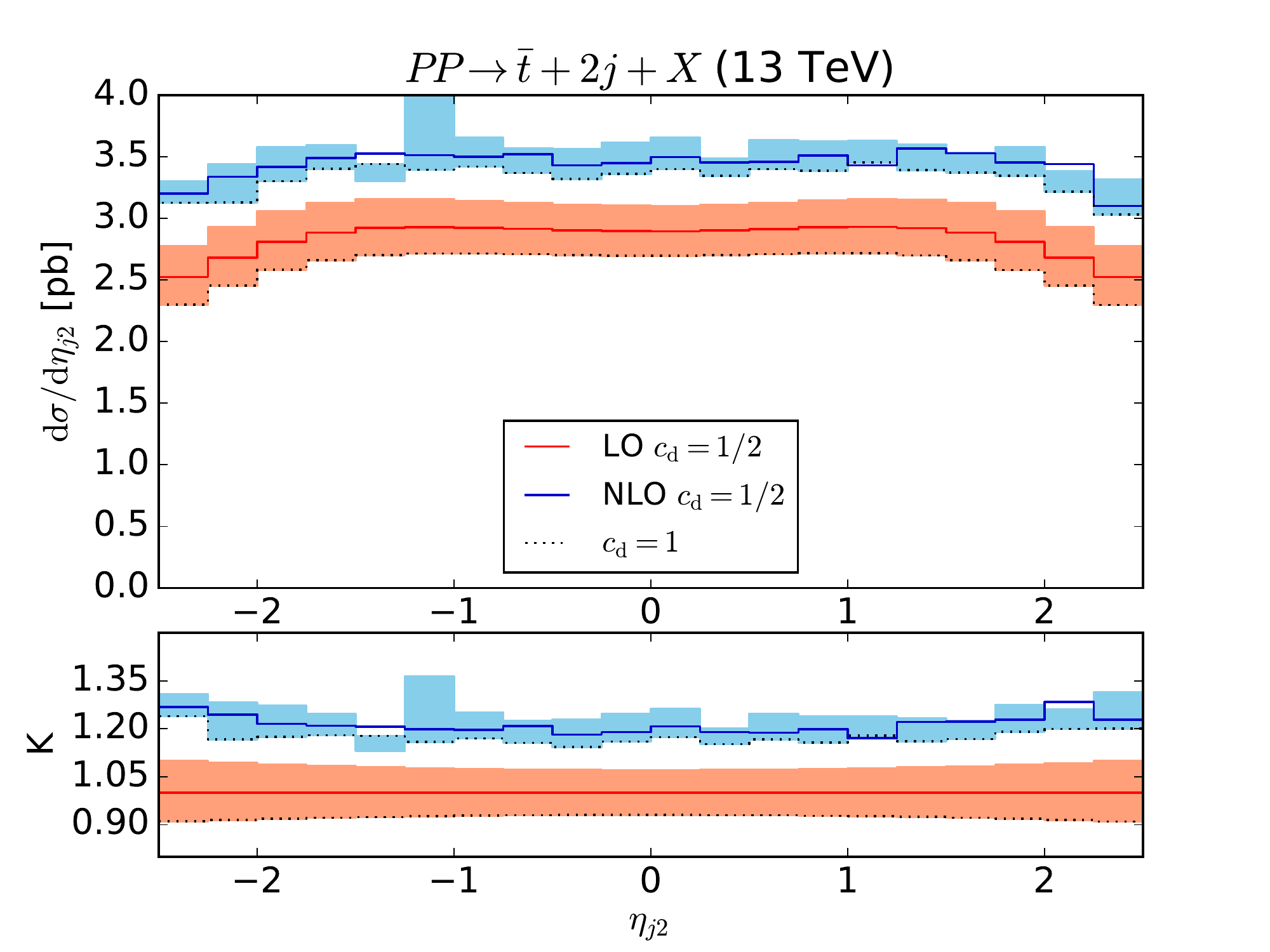}
  \end{tabular}
  \caption{Same as \fig{fig:dist_y_eta} but for anti top-quark
    production.   \label{fig:dist_y_eta_atop}}
\end{figure}

\begin{figure}[htbp]
  \begin{center}
    \includegraphics[width=0.6\textwidth]{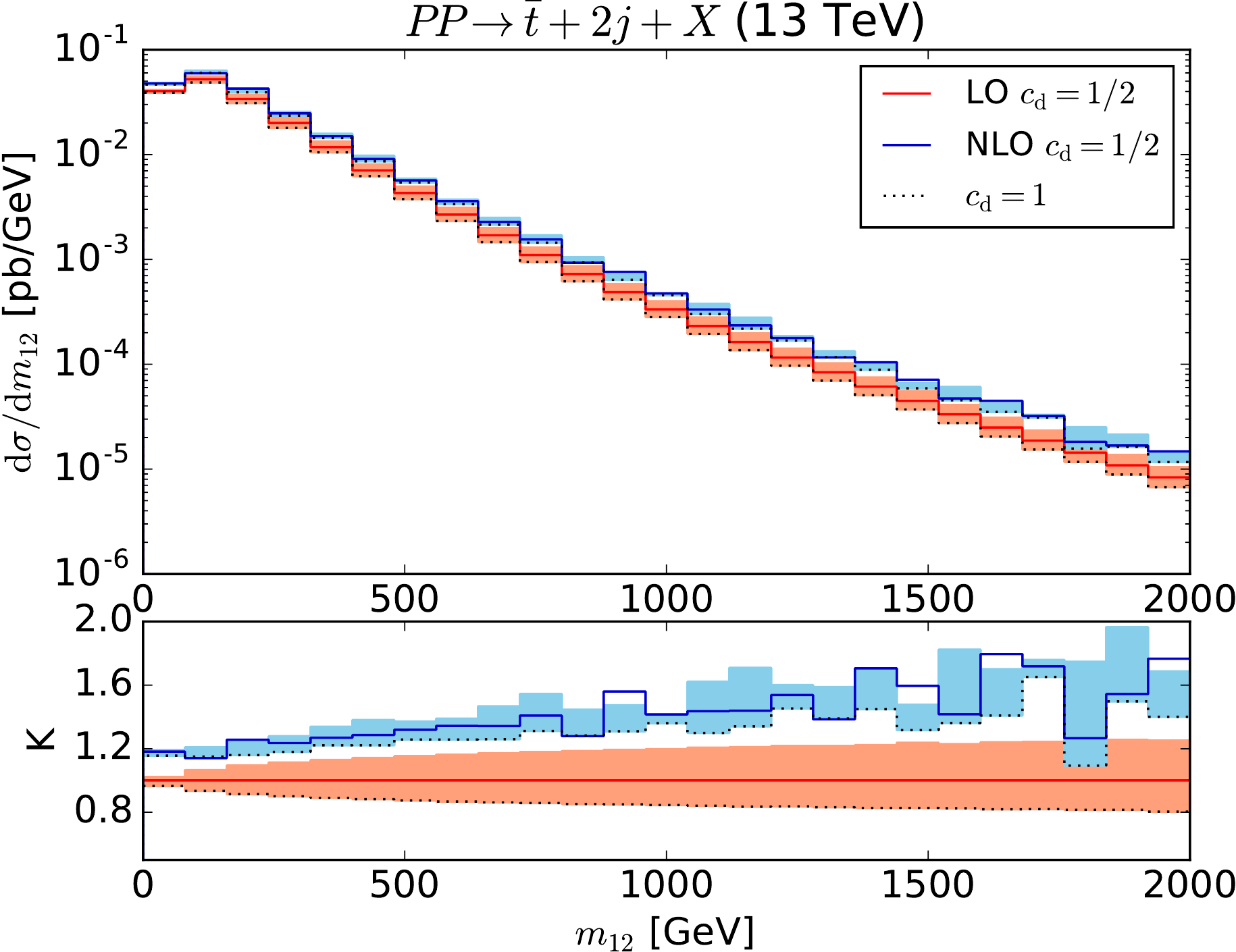}    
  \caption{Same as \fig{fig:m_jj} but for anti top-quark
    production. \label{fig:m_jj_atop}}
  \end{center}
\end{figure}

\begin{figure}[htbp]
  \begin{center}
    \includegraphics[width=0.6\textwidth]{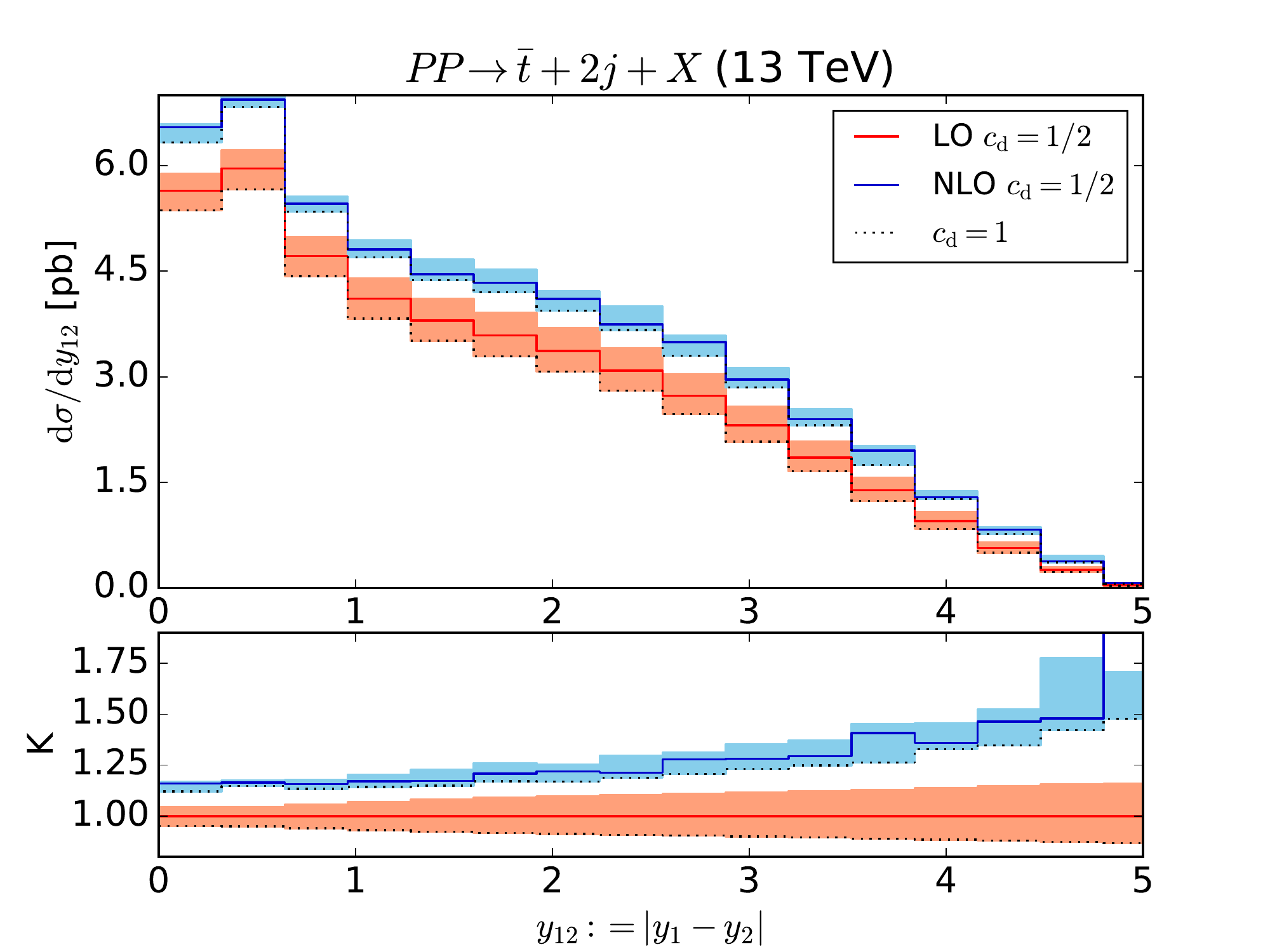}
    \includegraphics[width=0.6\textwidth]{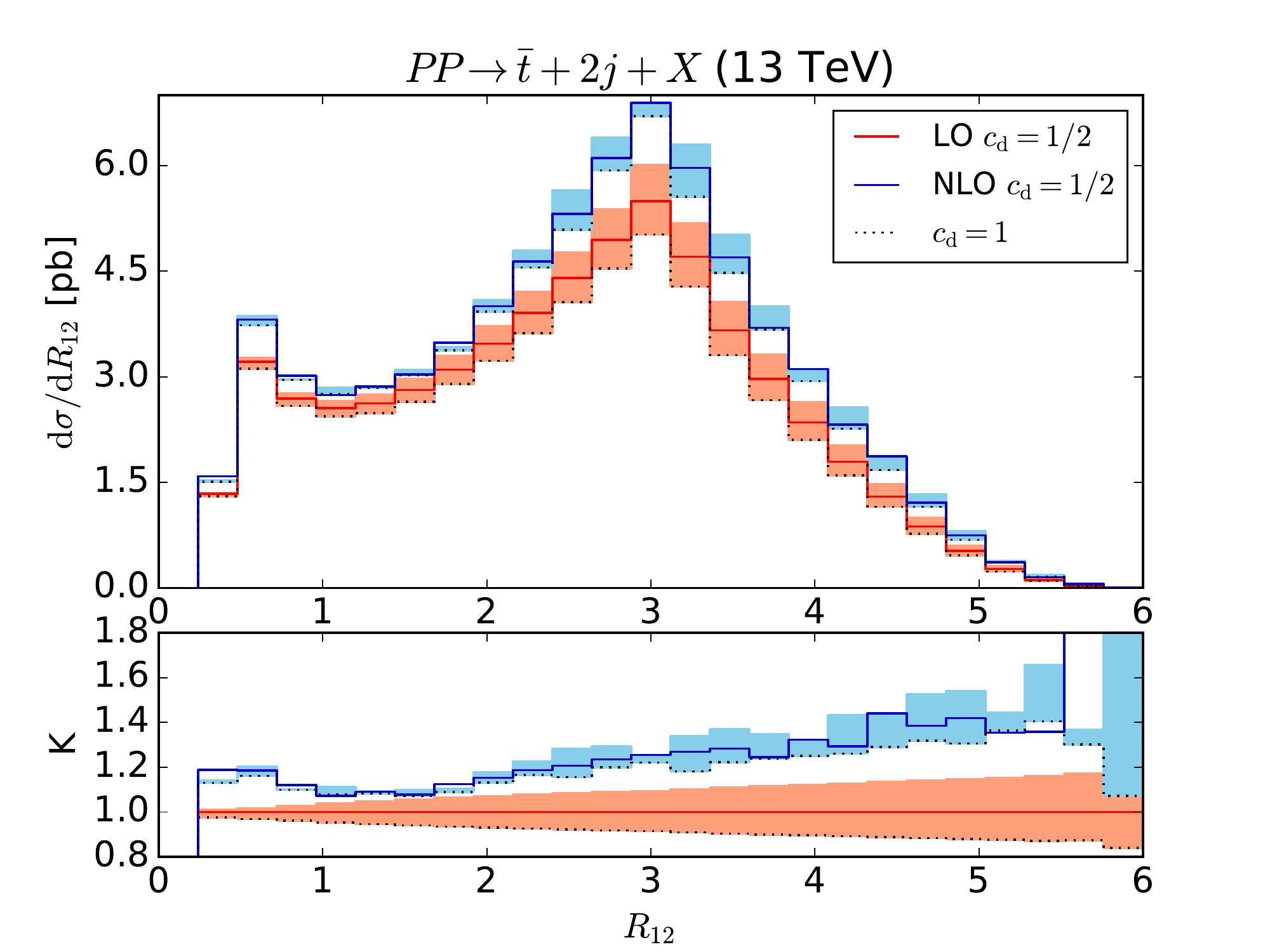}
  \end{center}
  \caption{Same as \fig{fig:dist_jj} but for anti top-quark
    production.\label{fig:dist_jj_atop}}
\end{figure}

In this appendix, results for anti top-quark production are provided.
Namely, the scale dependence is shown in
\fig{fig:scaledep_common-atop}, the transverse momentum distributions
of the anti top-quark and of the two leading jets are displayed in
\fig{fig:dist_pT_atop}, the rapidity distribution of the anti
top-quark and the pseudo-rapidity distributions of the two leading
jets are in \fig{fig:dist_y_eta_atop}, the invariant mass distribution
of the two leading jets is shown in \fig{fig:m_jj_atop}, and finally
the distributions of the rapidity and $R$ separation between the two
leading jets as defined in \eq{eq:Delta_y} and \eq{eq:Delta_R},
respectively, are presented in \fig{fig:dist_jj_atop}.

\clearpage
\newpage

\bibliographystyle{JHEP}
\bibliography{main}

\end{document}